\documentclass[aip,reprint]{revtex4-2}
\usepackage{gensymb}
\usepackage{mhchem}
\usepackage{placeins}
\usepackage{graphicx} %

\begin{document}
\title{Nanoconfined superionic water is a molecular superionic}

\author{Samuel W. Coles}
\email{swc46@cam.ac.uk}
\affiliation{\mbox{Yusuf Hamied Department of Chemistry, University of Cambridge, Lensfield Road, Cambridge, CB2 1EW, UK}}
\affiliation{\mbox{Lennard-Jones Centre, University of Cambridge, Trinity Ln, Cambridge, CB2 1TN, UK}}

\author{Amir Hajibabaei}
\affiliation{\mbox{Yusuf Hamied Department of Chemistry, University of Cambridge, Lensfield Road, Cambridge, CB2 1EW, UK}}
\affiliation{\mbox{Lennard-Jones Centre, University of Cambridge, Trinity Ln, Cambridge, CB2 1TN, UK}}

\author{Venkat Kapil}
\affiliation{\mbox{Department of Physics and Astronomy, University College London, 7-19 Gordon St, London WC1H 0AH, UK}}
\affiliation{\mbox{Thomas Young Centre and London Centre for Nanotechnology, 9 Gordon St, London WC1H 0AH, UK}}

\author{Xavier R. Advincula}
\affiliation{\mbox{Yusuf Hamied Department of Chemistry, University of Cambridge, Lensfield Road, Cambridge, CB2 1EW, UK}}
\affiliation{\mbox{Lennard-Jones Centre, University of Cambridge, Trinity Ln, Cambridge, CB2 1TN, UK}}
\affiliation{\mbox{Cavendish Laboratory, Department of Physics, University of Cambridge, Cambridge, CB3 0HE, UK}}

\author{Christoph Schran}
\affiliation{\mbox{Cavendish Laboratory, Department of Physics, University of Cambridge, Cambridge, CB3 0HE, UK}}
\affiliation{\mbox{Lennard-Jones Centre, University of Cambridge, Trinity Ln, Cambridge, CB2 1TN, UK}}

\author{Stephen J. Cox}
\affiliation{\mbox{Department of Chemistry, Durham University, South Road, Durham, DH1 3LE, UK}}

\author{Angelos Michaelides}
\email{am452@cam.ac.uk}

\affiliation{\mbox{Yusuf Hamied Department of Chemistry, University of Cambridge, Lensfield Road, Cambridge, CB2 1EW, UK}}
\affiliation{\mbox{Lennard-Jones Centre, University of Cambridge, Trinity Ln, Cambridge, CB2 1TN, UK}}

\date{\today}

\begin{abstract}
Superionic ice, where water molecules dissociate into a lattice of oxygen ions and a rapidly diffusing “gas” of protons, represents an exotic state of matter with broad implications for planetary interiors and energy applications. Recently a nanoconfined superionic state of water has been predicted which, in sharp contrast to bulk ice, is comprised of intact water molecules. Here we apply machine learning and electronic structure simulations to establish how nanoconfined water can be both molecular and superionic. We also explore what insights this material offers for superionic materials and behavior more generally. Similar to bulk ice and other superionic materials, nanoconfined water conducts via concerted chain-like proton migrations which cause the rapid propagation of defects. However, unlike other molecular phases of water, its exceptional conductivity arises from the activation of the Grotthuss mechanism by: (i) low barriers to proton transfer; and (ii) a flexible hydrogen bonded network. We propose that these are two key characteristics of fast ionic conduction in molecular superionics. The insights obtained here establish design principles for the discovery of other molecular superionic materials, with potential applications in energy storage and beyond.
\end{abstract}
\maketitle

Solid-state ionics, the study of solid electrolytes, has its origins in Faraday’s discovery of the superionic phase of silver sulfide in the 1840s\,\cite{Funke2013,Faraday_1843}. Faraday observed a phase transition on heating silver sulfide. The presence of a phase transition between two solid-state phases at elevated temperature was unsurprising. However, it was unexpected that the diffusion of silver ions in this higher-temperature solid phase exhibited an ionic conductivity equivalent to that of a liquid electrolyte. Later, scientists, drawing inspiration from electronic superconductivity, would term this phenomenon superionicity\,\cite{RICE1972294}.  Superionic behavior in chemically simple materials such as \ce{AgI} and \ce{PbF2} occurs at impractically high temperatures. However, recently designed materials based on lithium lanthanum zirconium oxide (LLZO)\,\cite{Thangadurai_2003,Murugan_2007}, lithium germanium thiophosphate (LGPS)\,\cite{Kamaya_2011}, and superionic polymers\,\cite{Paren_2022,Wang_2015} are currently being adopted for use in electrochemical devices.

Superionic water ice is one of the most widely studied and scientifically fascinating superionic materials. It is traditionally comprised of fully dissociated water molecules involving a “gas” of rapidly diffusing protons within a relatively stationary lattice of oxygen ions \,\cite{Millot_2018,Demontis_1988,Prakapenka_2021,Cheng_2021,Goldman_2005}. However, bulk superionic water phases (such as ice XVIII) 
form at the exceptional temperatures and pressures found in the cores of e.g. Uranus and Neptune ($ \sim 2000$ K, $ \sim 55$ GPa\,\cite{Demontis_1988,Prakapenka_2021,Cheng_2021,Goldman_2005,Cheng_2021,Matusalem_2022,Millot_2018}). 
Recently, a superionic state of water has been predicted to form at milder conditions when it is confined within sub-nanometer-wide slit pores \,\cite{Kapil_2022}.
This nano-confined superionic water is predicted to form at as low as 400 K and near the low GPa pressures naturally created within van der Waals bonded materials such as graphene\,\cite{Algara_Siller_2015}. Similar behavior has also been observed in recent experiments of nanoconfined water in slit pores, where there is evidence of exceptionally high ionic conductivity\,\cite{wang2025inplane}. A superionic state of water at or near room temperature is a tantalizing prospect scientifically and technologically.

In contrast to bulk superionic water and conventional superionic materials, simulations suggest that nano-confined water is comprised of intact water molecules\,\cite{Kapil_2022,Jiang_2024,ravindra2024nuclearquantumeffectsinduce}. 
However, as yet, there has been no comparison of the diffusive mechanism of this molecular system to more conventional superionic materials. In addition, it is unclear why it forms nor what “rules” govern the formation of superionic states under nanoconfinement.
In this paper, we use the results of machine learning molecular dynamics simulations and electronic structure calculations to reveal the chemical and physical nature of superionicity under nanoconfinement and to define the rules which govern its function and behavior. 

We confirm that nanoconfined water has a quasi-crystalline hexatic molecular structure, in sharp contrast to bulk superionic water\,\cite{Kapil_2022}, which is better described as a giant inorganic structure. Despite this difference, just as in conventional superionics, nanoconfined superionic water conducts via the rapid propagation of defects with a mechanism based on diffusive chains. In the case of nanoconfined superionic water, these defects are hydroxide and hydronium ions, which are the analogues of charge carrier vacancies and interstitial defects in conventional superionics.  These defect ions can form in this neutral system via autoionization due to the short separation between oxygen atoms and exist at exceptionally high concentrations compared to bulk aqueous systems. This means that this system is both molecular and superionic and can be described as a molecular superionic. We show that the high ionic conductivity in nanoconfinement, arises from facile proton transfer and hydrogen bond rearrangement\,\cite{Laage_2006,Hassanali_2013}, which are caused by short distances between oxygen atoms and dangling hydrogen bonds respectively\,\cite{Bernal_1933,Das_2024}. Finally, we consider how these insights can accelerate the discovery of other molecular superionic materials.

\section*{Results and discussion}
We begin by building on the work performed in prior studies characterizing the structures of nanoconfined\,\cite{Kapil_2022,ravindra2024nuclearquantumeffectsinduce,Jiang_2024} and bulk superionic water\,\cite{Goldman_2005,Sun_2015,Cheng_2021,Millot_2018}. In Figure 1 we explore the sharp contrast in the structure between nanoconfined superionic water and bulk superionic water.

\begin{figure*}
    \centering
    \includegraphics[width=0.7\textwidth]{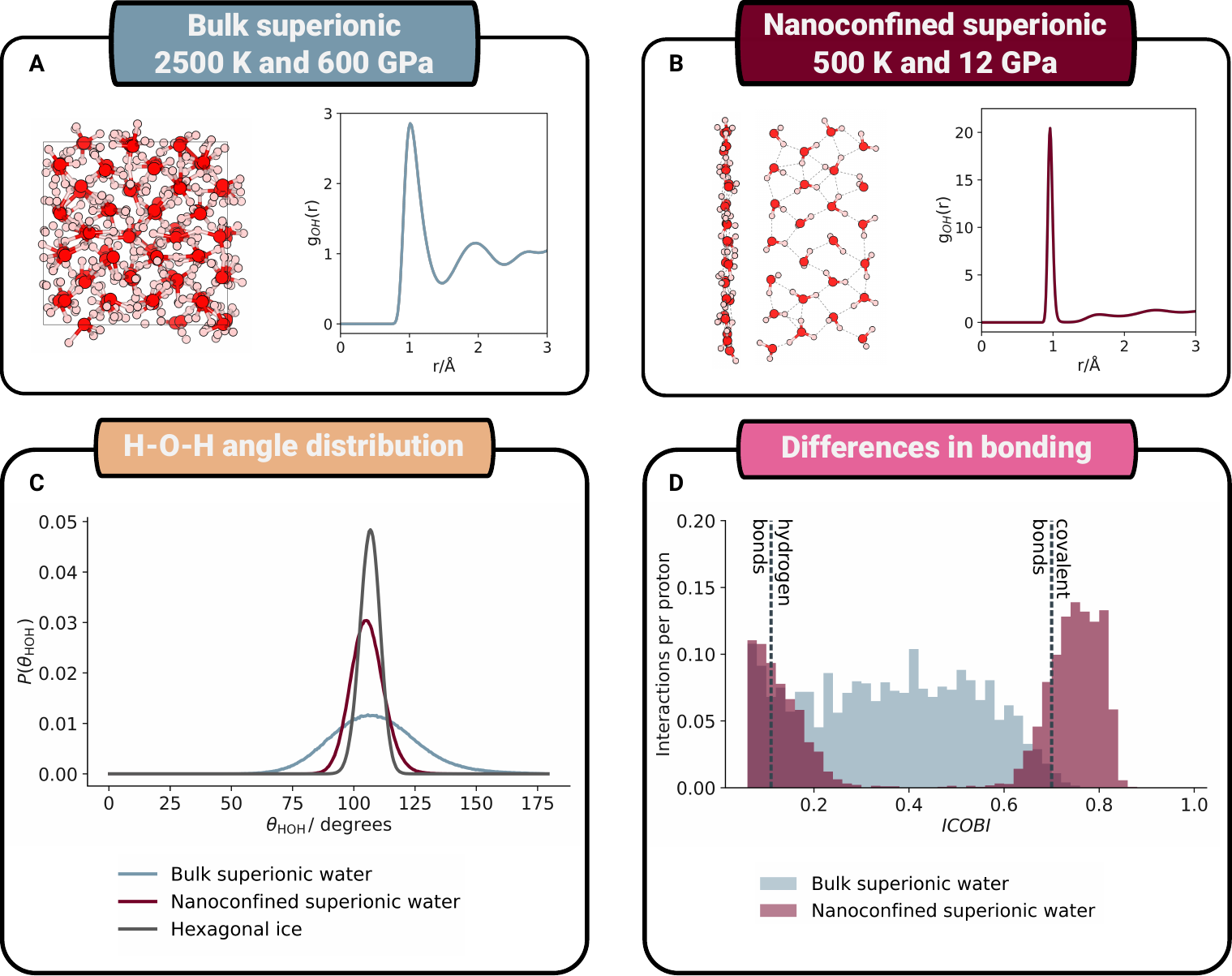}
    \caption{Bulk superionic water (panel A) is not molecular but nanoconfined superionic water (panel B) is. (A)--(B) Representative structural snapshots of the two phases (left) and plots of $g_\text{OH}(r)$ (right). In the case of the nanoconfined system both side and top views are shown of a selected portion of the unit cell. (C) Plots of the hydrogen-oxygen-hydrogen bond angles in the two phases as well as hexagonal ice at 250 K and 0.1 MPa. (D) Comparison of the chemical bonding in the two systems using an electronic structure descriptor, specifically the ICOBI index (see text for further details). The dotted lines on this panel reflect the average ICOBI values for the hydrogen bonds and covalent bonds in bulk hexagonal ice.}
    \label{fig:superionic_water}
\end{figure*}

We look first at the structure of (bcc-ordered) bulk superionic water (at 2500 K and 600 GPa (the area on the phase diagram where it exists)) in Fig. 1a. In the snapshot shown on the left we see the structure has a clear periodic arrangement of both oxygen and hydrogen ions, with protons normally centered on the interstices in the periodic lattice. The radial distribution function (RDF) between protons and oxygens in this system, which is presented on the right, has a form comprising a broad first peak with a magnitude similar to an ion pair in a molten salt\,\cite{Morgan_2004,Simoes_Santos_2024}, followed by an oscillatory decay to a value of 1 with a periodicity derived from the periodic ordering in the crystalline lattice. This form is characteristic of that between mobile and framework ions in a conventional superionic material (for instance AgI (an exemplar superionic material) shown in supporting information Fig. S6)\,\cite{Hull_2004}. Though differences exist between a conventional superionic and the bulk superionic water (for instance in the maximum value of ICOBI) they can both be described as ionically conductive inorganic crystals. This leads us to describe bulk superionic water as a conventional superionic material.

In contrast, if we consider these same descriptors for nanoconfined superionic water (Fig. 1b) the structure is clearly molecular. First, it is immediately clear on inspection of the structural snapshot that its constituent units are water molecules. A point that also leads to the sharp, narrow form of the first peak in $g_{\text{OH}}(r)$ caused by the strong molecular covalent bonds between oxygen and hydrogen. 

The difference in nature between the two systems is made clearer still if we consider the H-O-H bond angle in the two superionic systems and hexagonal ice (0.1 MPa, 250 K) shown in Fig. 1c. In the case of nanoconfined superionic water, the bond angle distribution is narrow, similar to that of hexagonal ice and centered on 106.3$\degree$. This is characteristic of the strong constraint placed on the bond angle in molecular water. For bulk superionic water, however, the distribution is different with a much broader range of bond angles being observed. This relative lack of constraint emerges from the non-molecular nature of the bonding in the phase, where the arrangement of protons is mainly templated by the geometry of the oxygen lattice as opposed to being driven by molecular geometry.

We have shown that the bulk and nano-confined superionic water have fundamentally different structures. However, it is not clear if this difference is correlated with an underlying difference in chemical bonding. To explore this we use density functional theory (DFT) to compute the electronic structure of each system and we then analyze this electronic structure with an approach known as the Integrated Crystal Orbital Bond Index (ICOBI)\,\cite{M_ller_2021}. ICOBI is one of many quantum chemical approaches for obtaining qualitative insights into bonding and is particularly useful when trying to distinguish ionic from covalent bonding. The value of ICOBI is a proxy for the covalency of a bond, with 1 being representative of a fully covalent interaction and 0 for an interaction where no covalency is observed (either due to a purely ionic interaction or when atoms do not interact at all). The main result from this analysis is shown in Fig. 1d, where we plot the distribution in ICOBI values for nanoconfined and bulk superionic water. The distribution for nanoconfined water consists of two peaks, which align with the values of the covalent and hydrogen bonds in hexagonal ice (shown as dotted lines). This is indicative of the molecular nature of nanoconfined superionic water. The plot for bulk superionic water is strikingly different. There is a broad range of ICOBI values continuously from 0.7 to 0, reflective of a system comprised of thermally fluctuating protons located primarily on interstitial sites in the bcc lattice. The form of this distribution is similar to that for the conventional superionic material silver iodide, which is shown in the supporting information (Fig. S5). Therefore, we can see that bulk superionic water is best understood as adopting an inorganic crystalline structure, with a degree of covalent character\,\cite{Goldman_2005,Sun_2015}. This, it should be noted, is not universal for all planetary superionic ices. For instance, at substantially lower temperatures (approximately 700 K with a pressure of 70 GPa), ammonia has been predicted to form a superionic state consisting primarily of molecular NH$_2$$^-$ and NH$_4$$^+$ ions\,\cite{Ninet_2012,Pickard_2008}. This superionic ammonia ice and nanoconfined superionic water are both superionic and molecular and therefore represents a distinct class of superionic materials; we term this class molecular superionics. 

\begin{figure*}
    \centering
    \includegraphics[width=0.7\textwidth]{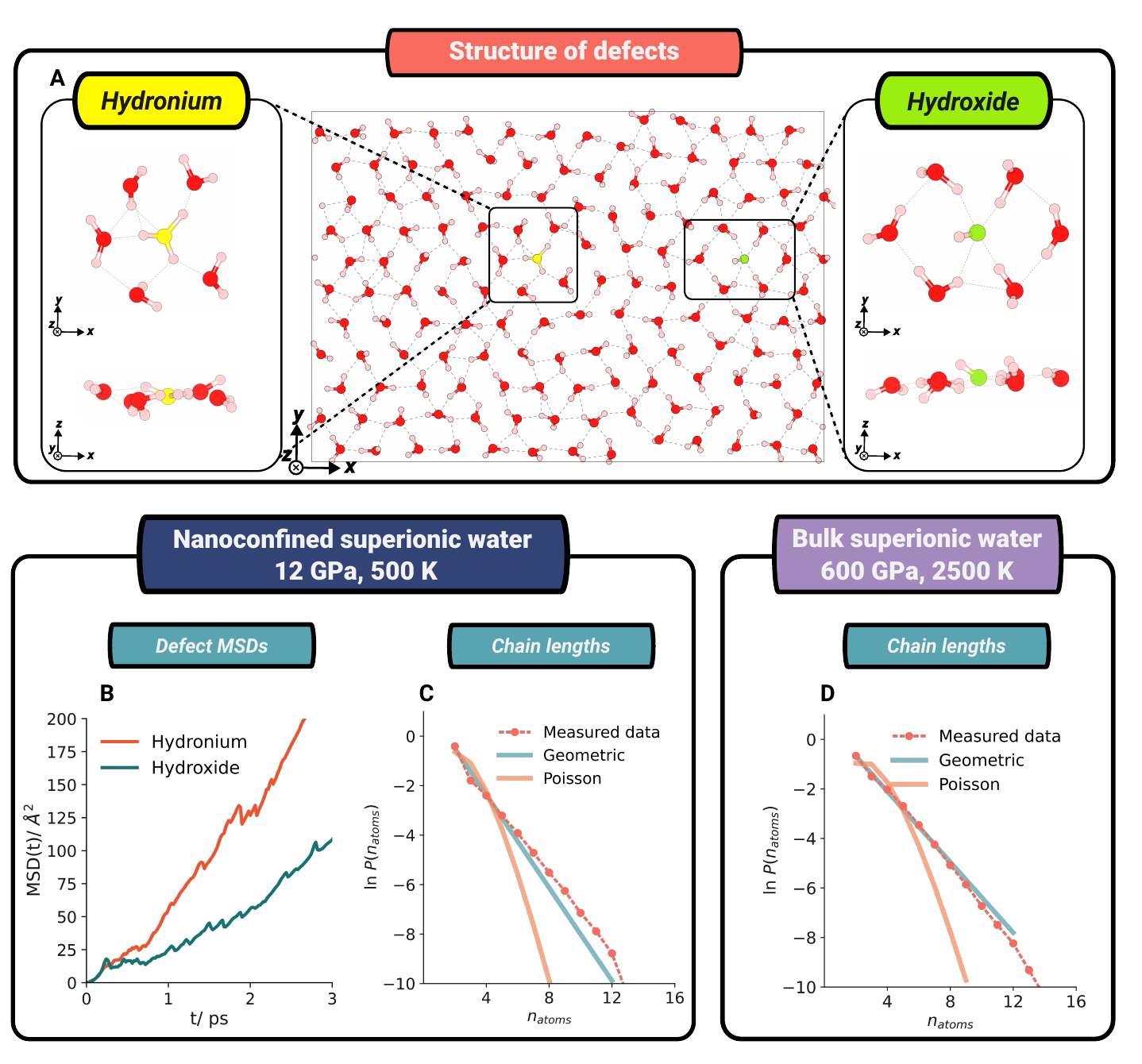}
    \caption{Both nanoconfined superionic water and bulk superionic water undergo chain-like defect-based diffusion.  (A) Snapshot of a nanoconfined system with both a hydroxide ion (oxygen atom is shown in green) and a hydronium ion (oxygen atom is shown in yellow) with zoomed in snapshots of the solvated structures of these ions. (B) MSD plots of the defects. (C)-(D) Probability of diffusive chains consisting of a certain number of hydrogen atoms in nanoconfined and bulk superionic water, respectively.}
    \label{fig:diffusion}
\end{figure*}

Materials have conventionally been considered superionic if they have a conductivity that exceeds 0.1 S/cm\,\cite{Kreuer_1996}. In addition, in the broader field of solid-state ionics, criteria for diffusive mechanisms in superionic materials have been proposed\,\cite{Hull_2004,Wood_2021,Catlow_1990}. Specifically, it has been suggested that in superionic materials diffusion occurs: (i) via defects; and (ii) correlated ion migrations. We now explore whether nanoconfined superionic water meets these descriptors. 

The conduction mechanisms in conventional superionic conductors often occur through the propagation of defects, specifically either interstitial ions or charge carrier vacancies\,\cite{Catlow_1990,Morgan_2014}.  During superionic conduction, these defects propagate through the system rapidly at a rate far faster than that of the individual charge carriers. In nanoconfined superionic water, the defects are solvated hydroxide and hydronium ions. These are shown in Fig. 2a and are direct analogs of the vacancies and interstitials that constitute the so-called Frenkel pairs in conventional superionic materials\,\cite{Joos_2025}. These defect pairs form spontaneously by the transfer of a proton from one water molecule to another. In some cases, the proton immediately transfers back leading to the annihilation of these defects. However, when this does not happen, they propagate through space via the Grotthuss mechanism\,\cite{de1805moire,Marx_2006}. This causes them to become separated like they are in the snapshot in Fig. 2a. Eventually, each defect will meet another alternatively charged defect and, if proton transfer occurs, annihilate. The average mean squared displacement of each defect species in Fig. 2b shows them both to be diffusive, with the hydronium defects diffusing faster than the hydroxide. The faster hydronium defect diffusion observed here is consistent with simulations of dilute defects in bulk water\,\cite{Chen_2018}. 

In bulk superionic water, the defect chemistry is based on conventional crystalline interstitials and vacancies, not hydroxide and hydronium ions. It is not appropriate to describe the oxygens as hydroxides and hydroniums due to the bonding of protons to multiple oxygen atoms. When we attempt to apply this nomenclature (Supporting Information, Fig. S10) some oxygens are arbitrarily assigned as \ce{O2-} and \ce{H4O2+}. In fact, the defect chemistry we observe in nanoconfined superionic water is far more reminiscent of conduction in acids and bases\,\cite{Tuckerman_2006,Marx_1999}. With the exceptionally low pKw of nanoconfined superionic water – ranging from 3 to 1.5 with temperature (discussed in Supporting Information section S4) – signifying concurrent concentrations of hydroxide and hydronium ions equivalent to strong bases and acids, respectively.

Let us consider now correlated ion migrations, as opposed to isolated hoping events consisting of the movement of a single ion\,\cite{Wood_2021,Burbano_2016,Morgan_2021}. In Fig. 2 we report the probability that a diffusive chain of proton hops in nanoconfined superionic water (panel c) and bulk superionic water (panel d) consists of a certain number of atoms. If the diffusive events were independent of one another and chains were forming via uncorrelated hops, we would expect these logarithmic graphs to follow Poisson distributions. It is clear, however, that they both have a geometric (straight line) form, indicating the presence of a chain-like diffusion mechanism comprising correlated hops. This chain-like diffusion is also observed for silver iodide (Supporting Information, Fig. S7) and multiple other superionic materials\,\cite{Burbano_2016,Morgan_2021,Annamareddy_2017}. Thus, like other superionic materials, nanoconfined water undergoes correlated ion motion. In contrast, however, the system remains molecular and the motion Grotthuss-like. It is therefore clear that nanoconfined superionic water is both intrinsically superionic and intrinsically different from other superionic materials. This further motivates our classification of it as a molecular superionic.

The prominence of the chain-like Grotthuss mechanism in nanoconfined superionic water’s diffusion mechanism may at first seem unsurprising, since this is the conventional mechanism of proton conduction in molecular water-based systems. 
However, there is – as we would expect from our structural analysis – an alternative chain-like mechanism of proton conduction in the non-molecular bulk superionic ice. 
In fact, as superionicity is not observed in any molecular water ices\,\cite{Futera_2020,Noguchi_2016} there must be a specific feature of nanoconfined superionic water – as well as the conductive molecular ammonia ices –  which allows Grotthuss to drive superionic conduction. 
Drawing on the commonality with conventional superionic conductors, we can propose two key features that are necessary for a material to be conductive. First, the barrier for charge carrier motion is sufficiently close to thermal energy that charge carriers can migrate\,\cite{Wood_2021}. Second, there are diffusive pathways in the system that allow an ion to move in a direction other than back and forth between neighboring sites\,\cite{Burbano_2016} (i.e. to do more than rattle within a single O-O pair). In Fig. 3. we explore how both criteria are met in nanoconfined superionic water in contrast with other neutral molecular water-based systems.

First, we investigate the free energy barrier to proton transfer for each system. These are shown in the left panels of Figs. 3c-e, where we plot a 2-dimensional free energy surface of oxygen-oxygen separation ($r_{OO}$) versus proton distance from the oxygen-oxygen midpoint ($\delta$). 
This is a widely used coordinate system for exploring proton transfer in aqueous systems\,\cite{Chen_2018,Marx_1999,Li_2010}, with the midpoint ($\delta=0$) representing the transition state for proton transfer (See Methods for further details). 
In the pressurized water plot (Fig. 3c left) there is a considerable barrier to proton transfer. 
%
%
In nanoconfined superionic water, the barrier is decreased as shown by the substantially lower saddle point in the free energy plot in the left panel of Fig. 3e indicating the energetic favorability of proton transfer in this phase. 
This decrease in barrier height arises from a pressure-driven decrease in separation between oxygens. In Fig. 3b we observe this decrease in the separation between oxygens with pressure in nanoconfined superionic water.  However, we also observe that bulk pressurized water and ice VII do not reach similar separations in the same pressure range. This suggests that the close separation – and with it the decreased proton transfer barrier – emerges from the twin effects of pressure and a nanoconfinement-induced change in the relationship between separation and the applied pressure. 

However, enabling proton transfer – the motion of protons between neighboring molecules – is not sufficient to activate diffusion in a molecular water-based material. Ice VII can reach oxygen separations similar to nanoconfined water at pressures over 25 GPa. At these separations, there is a high degree of proton transfer – as we can observe in the free energy plot on the left of Fig. 3d (at elevated temperatures in excess of 500 K). 
Despite this ice VII remains an exceptionally poor conductor under these conditions, with a conductivity six orders of magnitude lower than nanoconfined superionic water\,\cite{Noguchi_2016}. The origin of this difference lies in whether there are diffusive pathways in the two systems. In an inorganic superionic conductor, each site within this system is connected to multiple others, which leads to extended interconnected diffusive pathways. In contrast, in a molecular water-based system, intermolecular proton transfer occurs along hydrogen bonds. Proton transfer alone, therefore, cannot lead to proton diffusion (the long-range mass transport of protons), because after a proton transfer event there is nowhere for a proton to go other than the direction it came. 
Drawing inspiration from studies of proton transfer in acidic systems\,\cite{Marx_2006,Tuckerman_2006,Hassanali_2013,Laage_2006} and the recent work by Gomez\,\textit{et al.}\,\cite{Gomez_2024}, we propose the 2 step mechanism for diffusion in nanoconfined superionic water shown in Fig. 3a. In this mechanism, proton transfer is followed by molecular rotation, leading to a change in hydrogen bonding, after which any additional intermolecular proton transfer can occur to a third oxygen molecule. Repetition of this process opens diffusive pathways in the system allowing proton diffusion to occur.

In the rightmost plot of panels of c, d and e of Fig. 3, we investigate if this two-step mechanism is possible in ice VII, pressurized water, and nanoconfined superionic water. Taking inspiration from prior studies of proton transfer\,\cite{Hassanali_2013,Chandra_2007} we define two correlation functions $C_{PT}(t)$  and $C_{HB}(t)$ (which are defined mathematically in Methods). $C_{PT}(t)$ has a value of 1 when a proton is the nearest neighbor to the same oxygen at time t as it was at time 0, and a value of 0 otherwise. It describes the rate of change in covalent bonding and can be viewed as a proxy for proton transfer. $C_{HB}(t)$ is 1 if the two nearest neighbor oxygens to each proton are the same at time t and time 0 and has a value 0 otherwise. It is a proxy for the rearrangement of the hydrogen bond network and does not change in the event of proton transfer (a rattling event). The proton transfer correlation function ($C_{PT}(t)$) can be related to the previously discussed proton transfer-based free energy plots on the left side of Figs. 3 c, d, and e.

\begin{figure*}
    \centering
    \includegraphics[width=0.7\textwidth]{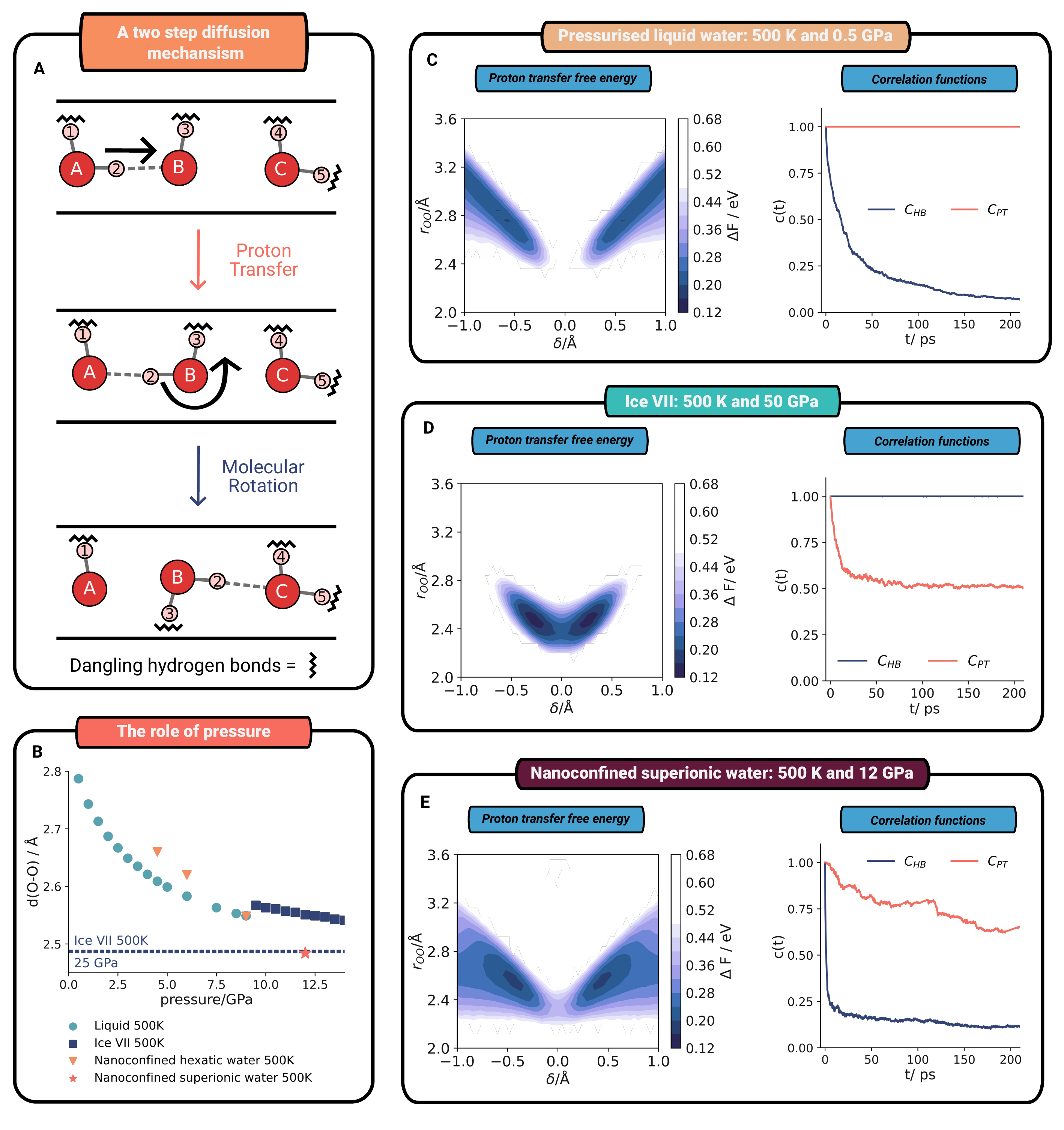}
    \caption{Closer separation of oxygen atoms and hydrogen bond network flexibility enables a two-step proton diffusion mechanism in nanoconfined superionic water. (A) A simplified scheme of the proton diffusion mechanism in nanoconfined superionic water. (B) Influence of pressure on the separations between oxygens in liquid water, ice VII, and nanoconfined superionic water. (C)-(E) Correlation functions and free energies of pressurized liquid water, ice VII, and nanoconfined superionic water. For each material, we present descriptors that show its ability to undergo the two-step diffusion process shown in (A). On the left-hand side panels, free energies for proton transfer are plotted as a function of oxygen-oxygen separation ($r_{\mathrm{OO}}$) and the location of the proton along the oxygen-oxygen axis ($\partial$). On the right-hand side panels, we present the correlation functions $C_{\mathrm{HB}}$ and $C_{\mathrm{PT}}$ (described in the main text and Methods) for each of the three phases.}
    \label{fig:correlations}
\end{figure*}

For pressurized liquid water at 500 K (Fig. 3c), we observe the familiar behavior of ambient liquid water. The hydrogen bond network, of course, rearranges rapidly which leads to the decay in $C_{HB}(t)$ in the correlation plot in Fig. 3c. Water molecules are, however, too well separated leading to a high barrier to proton transfer – as can be observed in the central plot. This leads $C_{PT}(t)$ to maintain a value of 1, yielding the familiar limited conductivity of pure water.

For ice VII (Fig. 3d), in the rightmost plot, we see prompt decay of $C_{PT}(t)$ to a value of 0.5 on a picosecond timescale, indicating the presence of proton transfer. This is consistent with the relatively low free energy barrier for proton transfer shown on the left of Fig. 3d. However, $C_{HB}(t)$ maintains a constant value of 1 throughout the length of this simulation. This lack of variance in $C_{HB}(t)$ arises from a persistent hydrogen bonding network, presumably because of the saturated nature of the hydrogen bonding network in ice VII. Thus, the full two-step mechanism of proton diffusion does not occur, and again this system is not conductive.

Nanoconfined superionic water exists within a “Goldilocks zone” between these two extremes. We observe decay in both $C_{PT}(t)$ and $C_{HB}(t)$ (Fig. 3e), revealing both proton transfer and hydrogen bond network rearrangement in the system and that the two-step mechanism of proton diffusion is viable. We have already noted the close separation of oxygens in nanoconfined superionic water, which enables proton transfer to occur. We suggest the reordering of the hydrogen bonding network, arises from the presence of dangling hydrogen bonds induced by nanoconfinement\,\cite{Ravindra_2024}. This means there are non-hydrogen-bonded hydrogen atoms within the system, which mainly, though not exclusively, sit above or below the plane of oxygen atoms. This makes the hydrogen bonding network more flexible, facilitating molecular rotation and proton conduction.

\section*{Discussion and Conclusions}

In summary, nanoconfined superionic water has a different structure from bulk superionic water and inorganic superionic materials. In contrast with these other superionic materials, it is composed of molecules. Despite these structural differences, it still conducts protons via the rapid propagation of defects — specifically hydroxide and hydronium ions — with a chain-like diffusion mechanism. We, therefore, conclude nanoconfined superionic water is a molecular superionic. 

The possibility of inducing superionicity in ice VII has previously been discussed\,\cite{Noguchi_2016,Futera_2020}. However, no molecular phase of pure water has been observed to exhibit superionicity under equilibrium conditions either experimentally or computationally\,\cite{Futera_2020}. 
We have shown nanoconfined superionic water’s unique conductivity is characterized by the presence of both proton transfer and a constantly rearranging hydrogen bond network. 
Proton transfer occurs primarily because of the closer separation between oxygen atoms at the elevated pressure at which the phase forms. 
The constant rearrangement of the hydrogen bond network is made possible by the increased flexibility arising from the lower number of hydrogen bonds formed by each water molecule, and the associated presence of dangling hydrogen bonds\,\cite{Das_2024}. 
In this water-based nanoconfined superionic material, this conductivity occurs via the Grotthuss mechanism. 
The previously mentioned conductive molecular ammonia phase, which exists at similar temperatures and pressures to ice VII\,\cite{Ninet_2012,Pickard_2008}, can be viewed as a bulk crystalline molecular superionic. 
The difference in conductivity between this phase and ice VII is in ammonia’s molecular structure. Each ammonia molecule can donate three hydrogen bonds but can only accept one. 
This means that in this dense ammonia-based molecular crystal, there will always be dangling hydrogen bonds in contrast to pristine water ice phases which have saturated hydrogen bond networks
(though a similar effect could be obtained by the wholesale incorporation of defects). 
This also allows these ammonia-based crystalline materials to behave as molecular superionics. 
Just as in nanoconfined superionic water, conduction in these crystals is correlated with large amounts of disproportionation (in this case, to form NH$_4$$^+$ and NH$_2$$^-$) and a huge increase in the rotation of individual molecules in this superionic phase compared to non-conductive phases\,\cite{Ninet_2012}. 
The similarities in behavior between this ammonia-based phase and nanoconfined superionic water suggest that molecular superionicity is, in general, characterized by the presence of both a low-charge carrier barrier and a hydrogen bonding network, which actively rearranges, enabling long-range diffusion of charge carriers.

Let us now discuss how molecular superionicity might be observed in other systems, particularly under conditions closer to room temperature and pressure. 
First, mixtures of hydrogen-bonding molecules are likely to be a promising avenue for further exploration. In these systems disorder, and possibly variations in pH, could introduce both the dangling hydrogen bonds and intrinsic defects necessary to induce molecular superionicity. 
Second, a similar strategy of materials discovery that was used to discover both nanoconfined superionic water\,\cite{Kapil_2022} and superionic molecular ammonia\,\cite{Pickard_2008,Ninet_2012} could be applied to predict superionic phases in nanoconfinement which form at conditions closer to ambient conditions. This could involve the exploration of different combinations of hydrogen bonded molecules and different confining materials. Such systems are promising because they are now readily amenable to experimental investigation and can potentially support superionic states 
at milder conditions. 

A third, and likely initially simpler route would be to induce molecular superionicity at interfaces or within porous frameworks at reduced or ambient pressure. Reactive interfaces can introduce dangling hydrogen bonds in the same manner as nanoconfinement, so it would only be necessary for the specific interface to template a reduction in the separation of oxygen atoms for both of our proposed criteria for molecular superionicity. Encouragingly, this templating effect has already been observed at the reactive interfaces of metals and oxides where water strongly absorbs at certain sites, creating interfacial regions composed of a mixture of water molecules and hydroxide ions\,\cite{Li_2010,Tocci_2014}. This third approach would create a localized superionicity, not a superionic bulk material. This would bring with it a fine control of localized conductivity. This suggests molecular superionicity could be a helpful tool in nanoscale devices and the control of sensitive surface-based aqueous chemical processes.  

Finally, we note on a broader point that the similarities and differences between conventional and molecular superionics described here are significant. In this paper, we have discussed three different definitions of superionicity. That all three of them apply to both conventional superionics\,\cite{Morgan_2014} and nanoconfined superionic water suggests that: high conductivity, chain-like diffusion, and the presence of highly diffusive defects can be viewed as transferable criteria of superionicity

\section*{Methods}

\subsection*{Potentials and dynamics}

We begin by discussing the potentials and settings used for the molecular dynamics simulations in this paper. Detailed descriptions of the specific systems modelled can be found in the Supporting Information.

\subsubsection*{Nanoconfined superionic water}

The methodology of this paper was built around machine-learned molecular dynamics of bulk and nanoconfined water. The machine learning potential used to model nanoconfined water was developed by Kapil \textit{et al.}\,\cite{Kapil_2022} and is a neural network potential fit to a training set of calculations at the revPBE0-D3 level of theory, based on the procedure described in Ref.\,\citenum{Schran_2021}. Data presented in this paper for nanoconfined superionic water is taken from simulations of the canonical ensemble with a timestep of 0.25 fs, with a Nosé-Hoover thermostat and damping parameter of 20 fs. This simulation is run at the average density from a prior calculation in an NP$_{operatorname{xy}}$T ensemble, where the barostat only acts in the xy dimension. As in the prior study by Kapil \textit{et al.} simulations were performed using i-PI 2.0\,\cite{Kapil_2019} with interatomic forces calculated using the n2p2 lammps library\,\cite{Singraber_2019}. As in previous studies of nanoconfined superionic water confinement is modelled implicitly using a Morse potential, fitted to recover the interactions obtained in quantum Monte Carlo simulations of water on graphene\,\cite{Kapil_2022}, with the forces arising from this potential being calculated using ASE\,\cite{Hjorth_Larsen_2017}.  This confining potential is oriented such that it is uniform in the xy plane. A full structural description of the system is provided in the Supporting Information (Fig. S1). Simulations were run to between 1 ns and 2 ns in length depending on temperature. Extensive validations of the neural network potential can be found in the original publication of Kapil \textit{et al.}\,\cite{Kapil_2022}.

\subsubsection*{Bulk ices: Ice VII and bulk superionic water}

Simulations of ice VII and bulk superionic water were performed using a neural network potential\,\cite{Behler_2007} previously used and parameterized by Cheng \textit{et al.}\,\cite{Cheng_2021} using a training set of DFT calculations at the PBE-D3 level of theory, which has been applied previously to the study of bulk superionic ices\,\cite{Cheng_2021} and high-pressure molecular ices\,\cite{Reinhardt_2022}. Simulations were run with the lammps molecular dynamics software\,\cite{Thompson_2022}, with the neural network potential implemented using the n2p2 lammps library\,\cite{Singraber_2019}. Simulations were run with a 0.25 fs timestep. A two-step procedure was followed with an initial isotropic NPT run of 1 ns at a target temperature being used to calculate the equilibrium volume of the cell with a further 1 ns NVT run performed to calculate structural details. All simulations were run with a temperature damping parameter of 20 fs and a pressure damping parameter of 200 fs within the Nosé-Hoover thermostat and barostat.

\subsection*{Electronic structure calculations, defect tracking and workflow}

DFT calculations were performed with VASP\,\cite{Kresse_1994,Kresse_1996,KRESSE199615} using projector augmented wave (PAW) potentials, the PBE exchange-correlation functional\,\cite{Perdew_1996} and a 700 eV plane wave cut-off. ICOBI values were calculated from single-point VASP calculations using the LOBSTER software\,\cite{Nelson_2020}, with the pbeVaspFit2015 basis set. The structures considered were taken from the neural network potential molecular dynamics simulations and for the data reported in Fig. 1 ten structures chosen at equal intervals throughout the trajectory from which the associated RDF was chosen were used.  

\subsection*{Analytical methods}

Identification of defects and assignment of protons to oxygen atoms in the nanoconfined system is performed via Voronoi tessellation. This is appropriate as the highly molecular nature of the system means most protons are closest to a single oxygen ion with which they are covalently bonded. The codes for this and other simple analyses use workflow built on the pymatgen\,\cite{Ong_2013}, ASE\,\cite{Hjorth_Larsen_2017}, vasppy\,\cite{Morgan_Vasppy}, and numpy\,\cite{Harris_2020} scientific python libraries. Nanoconfined radial distribution functions are subject to the same finite size correction described by Fong \textit{et al.}\,\cite{Fong_2024}.

\subsection*{Calculation of chain lengths}

The calculation of chain lengths follows the established methods used in the study of diffusive chains in the field of solid electrolytes\,\cite{Morgan_2021,Annamareddy_2017} based on historic studies of glasses\,\cite{Donati_1998}. Chains are identified on a pairwise basis where one atom – in our case limited to the hydrogen atoms in the system – has moved away from its initial site and been replaced by another. This is defined mathematically for atoms with positions $\mathbf{r}_i$ and $\mathbf{r}_j$ at time t as,

\begin{equation}
\min \left[ \left\| \mathbf{r}_i(t + \Delta t) - \mathbf{r}_j(t) \right\|, \left\| \mathbf{r}_j(t + \Delta t) - \mathbf{r}_i(t) \right\| \right] < \delta
\label{eq:chains}
\end{equation}

\noindent for a time window $\Delta t$ and a spatial cutoff $\delta$. For both systems we use $\Delta t$ values of 100 ps and a $\delta$ value of 1.75 \AA{}. The latter value is taken to be less than the separation between physical sites in each system, which is the analogue to the hard sphere radius in the simple glasses for which these methods were initially defined. The code used to calculate these chains is adapted from previous work\,\cite{morgan2020argyrodite}.

\subsection*{Correlation functions}

The correlation function $C_{PT}(t)$ is defined as,

\begin{equation}
C_{PT}(t) = \langle C_{ij}(0) C_{ij}(t) \rangle
\label{eq:PT}
\end{equation}

\noindent where $C_{ij}$ is 1 where an oxygen $i$ is the nearest neighbor to hydrogen atom $j$ (a proxy for a covalent bond between the two atoms) and 0 otherwise. This creates a correlation function for changes in covalent bonding. While $C_{HB}(t)$ is defined as,

\begin{equation}
\begin{split}
C_{HB}(t) =\ & \langle C_{ij}(0) C_{ij}(t)  H_{ik}(0) H_{ik}(t) \\
& + C_{ij}(0) C_{ik}(t) H_{ik}(0) H_{ij}(t) \rangle
\end{split}
\label{eq:HB}
\end{equation}

\noindent where $H_{ij}$ is 1 when oxygen i is next nearest neighbor to hydrogen atom $j$ and $0$ otherwise, and k is the index of an additional hydrogen atom. This equation can be viewed as a proxy for hydrogen bonding. Correlation functions were calculated using multiple time origins.

\subsection*{Proton transfer free energy}

In Fig. 3 we present plots of free energy of proton transfer: pressurized water, ice VII and nanoconfined superionic water to undergo proton transfer and molecular rotations (the two components of our two-step mechanism). For proton transfer we use a two dimensional plot of free energy related to two variables $\delta$ and $\boldsymbol{r}_{OO}$ which has been widely used in the literature\,\cite{Chen_2018,Marx_1999,Li_2010,Tocci_2014}. $\delta$, the proton transfer coordinate is defined for a proton located at $\boldsymbol{r}_{H}$ between two oxygen atoms located at $\boldsymbol{r}_{O_1}$ and $\boldsymbol{r}_{O_1}$ and is mathematically defined as,

\begin{equation}
    \delta =\sqrt{\left(\boldsymbol{r}_H-\boldsymbol{r}_{O_1}\right)^2}-\sqrt{\left(\boldsymbol{r}_H-\boldsymbol{r}_{O_2}\right)^2}
\label{eq:deltapes}
\end{equation}

\noindent  This means that $\delta$ has a value of 0 for a proton perfectly equidistant between two oxygens. The separation between the two oxygens  $\boldsymbol{r}_{OO}$ is calculated as,

\begin{equation}
\boldsymbol{r}_{O O}=\sqrt{\left(\boldsymbol{r}_{O_2}-\boldsymbol{r}_{O_1}\right)^2}
\label{eq:roopes}
\end{equation}

\noindent The free energy relative to these variables is then calculated as,

\begin{equation}
F\left(\partial, \boldsymbol{r}_{O O}\right)=-k_B \ln P\left(\partial, \boldsymbol{r}_{O O}\right)
\label{eq:fepes}
\end{equation}

\section*{Acknowledgements}

This work was supported by the European Union under the “n-AQUA” European Research Council project (Grant No. 101071937). We are grateful for computational support and resources from the UK Materials and Molecular Modeling Hub which is partially funded by EPSRC (Grant Nos. EP/P020194/1 and EP/T022213/1). Access for both the UK Materials and Molecular Modeling Hub and ARCHER2 were obtained via the UK Car-Parrinello consortium, funded by EPSRC grant reference EP/P022561/1. S.J.C. is a Royal Society University Research Fellow at Durham University (URF\textbackslash R1\textbackslash 211144).

\section*{References}
\begingroup
\renewcommand{\section}[2]{} 

\endgroup



%

\clearpage
\onecolumngrid

\begin{center}
\textbf{\large Supporting Information: Nanoconfined superionic water is a molecular superionic}\\[2ex]

Samuel W. Coles\textsuperscript{1,2,*},
Amir Hajibabaei\textsuperscript{1,2},
Venkat Kapil\textsuperscript{3,4},
Xavier R. Advincula\textsuperscript{1,2,5},
Christoph Schran\textsuperscript{2,5},
Stephen J. Cox\textsuperscript{6},
Angelos Michaelides\textsuperscript{1,2,$\dagger$}
\\[2ex]

\textsuperscript{1}Yusuf Hamied Department of Chemistry, University of Cambridge, Lensfield Road, Cambridge CB2 1EW, UK\\
\textsuperscript{2}Lennard-Jones Centre, University of Cambridge, Trinity Ln, Cambridge CB2 1TN, UK\\
\textsuperscript{3}Department of Physics and Astronomy, University College London, 7-19 Gordon St, London WC1H 0AH, UK\\
\textsuperscript{4}Thomas Young Centre and London Centre for Nanotechnology, 9 Gordon St, London WC1H 0AH, UK\\
\textsuperscript{5}Cavendish Laboratory, Department of Physics, University of Cambridge, Cambridge CB3 0HE, UK\\
\textsuperscript{6}Department of Chemistry, Durham University, South Road, Durham DH1 3LE, UK\\[2ex]

\textsuperscript{*}{swc46@cam.ac.uk}, \textsuperscript{$\dagger$}{am452@cam.ac.uk}

\end{center}
\vspace{1em}

\setcounter{figure}{0}           
\renewcommand{\thefigure}{S\arabic{figure}}  
\setcounter{table}{0}
\renewcommand{\thetable}{S\arabic{table}}

\setcounter{equation}{0}
\renewcommand{\theequation}{S\arabic{equation}}

\onecolumngrid
\section*{S1: Structural  description of studied systems}
\FloatBarrier
\subsection{Nanoconfined superionic water}
The setup used to study nanoconfined water in this paper is based on that used in Ref.~\cite{Kapil_2022}. All structures in studies of nanoconfined superionic water have the form shown in Fig.~S1. Water molecules occupy a periodic space in the $xy$ plane. In the $z$ direction, an additional Morse potential is added to simulate the confining effect of graphene in a nanoslit of roughly 5~\AA{} of the overall 28.33~\AA{} of space in the $z$ dimension. The remainder of space in the $z$ dimension functioning as a vacuum slab greater than the cutoff of the neural network potential to prevent interaction with periodic images in that dimension.

\begin{figure}[h]
    \centering
    \includegraphics[width=0.4\textwidth]{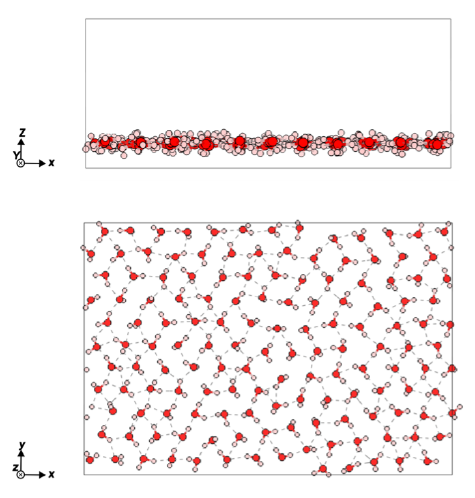}
    \caption{Structural plans of the nanoconfined system of 144 water molecules, where oxygen atoms are shown in red and hydrogens in pink. Water molecules are confined by an additional Morse potential in the $z$ dimension, which confines the water molecules to a roughly 5~\AA{} slit.}
\end{figure}

In general, simulations are performed in the NVT ensemble of 35.72~\AA{} in the $x$ dimension and 24.46~\AA{} in the $y$ dimension. The data from nanoconfined superionic water in Fig.~3B are generated from simulations previously run for Ref.~\cite{ravindra2024nuclearquantumeffectsinduce}. These simulations were run using an NP$_{xy}$T ensemble (with the $z$ dimension fixed with no change in the width of either the confining potential or vacuum slab).
\FloatBarrier
\subsection{Bulk superionic water}
\FloatBarrier

Simulations and calculations of bulk superionic water are performed using the structure in Fig.~S2. This structure has dimensions 9.72~\AA{} $\times$ 9.81~\AA{} $\times$ 9.81~\AA{} and contains 384 atoms. The small size of the cell was chosen so that ICOBI calculations and structural descriptors could be calculated from the same cell. A doubling of the cell size leads to negligible change in structure as evidenced by the similarities in the form of $g_{\operatorname{OH}}(r)$ in Fig. S3.
\begin{figure}[h!]
    \centering
    \includegraphics[width=0.4\textwidth]{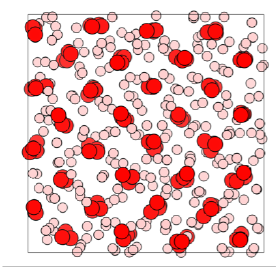}
    \caption{The structure of the bulk water used in the production of Fig.~1 in the main paper. The structure consists of a bcc oxygen lattice. In the initial structure, protons are placed in locations such that molecular units could reasonably form. In this structure, oxygens are shown in red and hydrogens in pink.}
\end{figure}

\begin{figure}[htbp]
    \centering
    \includegraphics[width=0.3\textwidth]{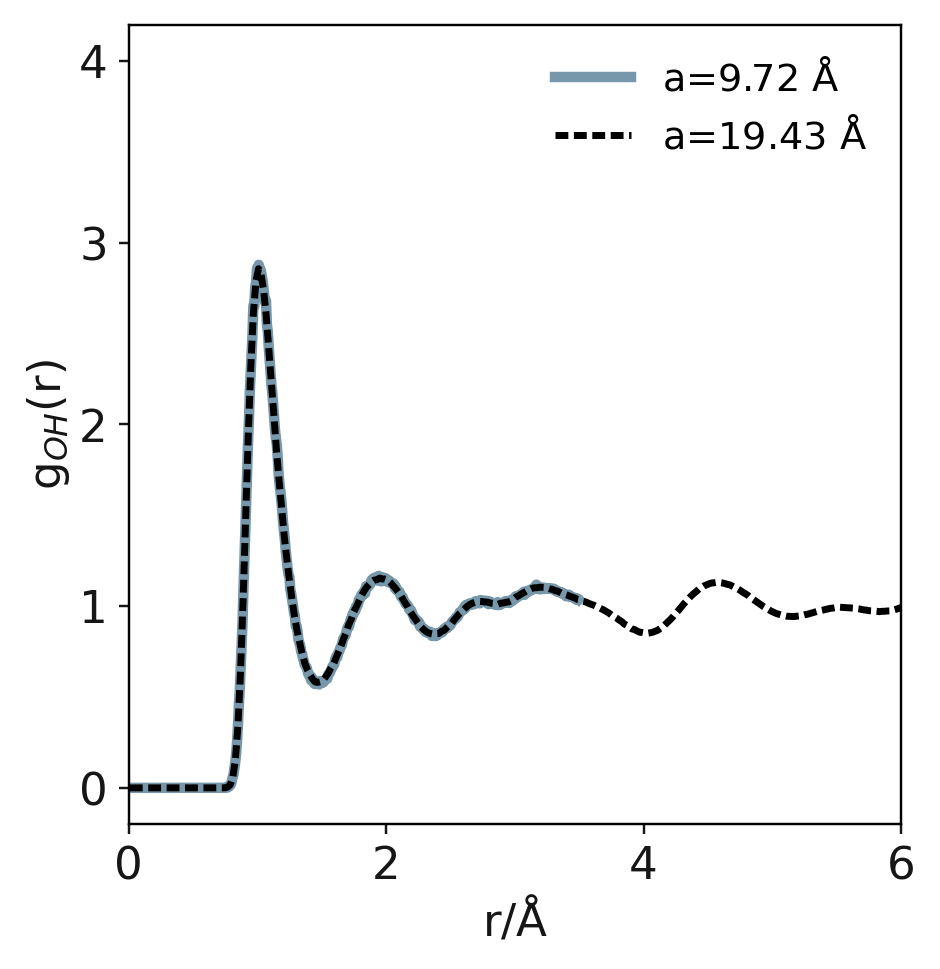}
    \caption{Reproduction of the radial distribution function of bulk water from Fig.~1 in the main paper, showing the same result for the small cell used in the main paper and a $2 \times 2 \times 2$ expansion of the primitive.}
\end{figure}

\FloatBarrier
\subsection{Ice VII and pressurized water}

Simulations and calculations of Ice VII were performed in a near-orthorhombic cell formed by a $6 \times 6 \times 4$ expansion of the ice VII primitive cell (as shown in Fig. S4). Simulations are performed in the NPT ensemble for the data points used to calculate the distances between oxygens. The calculations of correlation functions are performed in an NVT ensemble with a cell that has the same volume as the equilibrium volume from the 50~GPa run. This results in a cell with box side lengths 17.39~\AA{} $\times$ 17.38~\AA{} $\times$ 10.96~\AA{}.

\begin{figure}[h]
    \centering
    \includegraphics[width=0.4\textwidth]{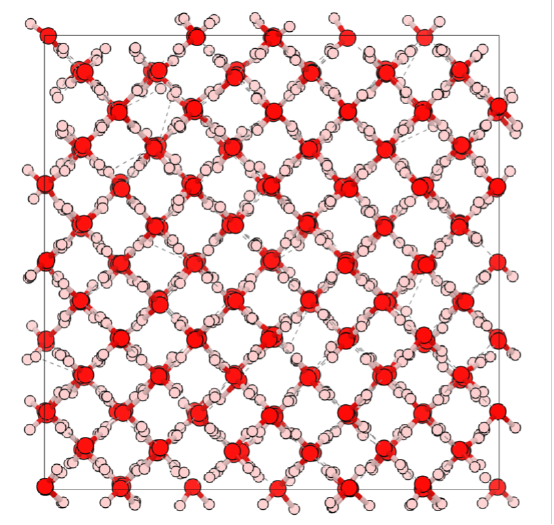}
    \caption{The structure of ice VII used in the production of the ice VII data points in Fig.~3 in the main paper. In this structure, oxygens are shown in red and hydrogens in pink.}
\end{figure}

This system is also the origin of the pressurized water systems. When ice VII is simulated at a temperature of 500~K and a pressure below 9.5~GPa, the system melts. The calculations for Fig.~3B are performed in the NPT ensemble with an anisotropic thermostat. The calculations for Fig.~3D are performed in the NVT ensemble in a cell with the average NPT volume; this cell has dimensions 24.39~\AA{} $\times$ 24.38~\AA{} $\times$ 15.37~\AA{}.

\FloatBarrier
\subsection{Superionic silver iodide}

\begin{figure}[h]
    \centering
    \includegraphics[width=.4\textwidth]{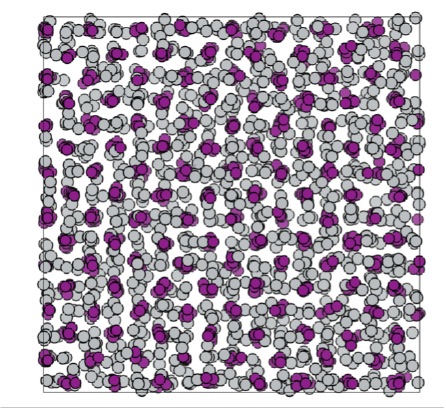}
    \caption{Structure of the silver iodide system used for the analyses in this supporting document, with silver shown in silver and iodine in purple. This structure is an 8$\times$8$\times$8 expansion of the primitive cell.}
\end{figure}

Simulations of silver iodide in section S3 were performed using the machine-learned MACE model\,\cite{batatia2022mace} trained on a set of PBE-D3 DFT calculations previously developed and validated by Hajibabaei \textit{et al.}\,\cite{Hajibabaei_2025} Simulations were performed using ASE in the NVT ensemble, with the volume set to match the equilibrium volume obtained from an NPT simulation. A Nosé-Hoover thermostat with a relaxation time of 1~ps was applied. A timestep of 1~fs was used due to the high mass of silver, and each simulation was run for 1~ns.

The structure of superionic silver iodide—shown in Fig.~S5—consists of 2048 atoms in an $8 \times 8 \times 8$ expansion of a primitive cell (40.62~\AA{} $\times$ 40.62~\AA{} $\times$ 40.62~\AA{}). For each temperature, the volume of the cell is taken from the equilibrium volume of a simulation run with a Nosé-Hoover barostat at a pressure of 1~bar. A smaller cell—a $4 \times 4 \times 4$ expansion of the primitive, containing 256 atoms—generated in the same way was used for the ICOBI calculations below.

\FloatBarrier
\section*{S2: Details of simulations and sampling used to make figures in the main text and in this supporting information.}

The actual dynamical simulations from which the analyses in the main paper and of silver iodide in section S3 are calculated are listed in Table S1.

\begin{table}[h!]
\centering
\renewcommand{\arraystretch}{1.3}
\begin{tabular}{|p{3.5cm}|p{2.2cm}|p{1.5cm}|p{2cm}|p{3cm}|p{3cm}|}
\hline
\textbf{System} & \textbf{Temperature} & \textbf{Length} & \textbf{Number of atoms} & \textbf{Figures Calculated} & \textbf{Figures Derived} \\
\hline
Nanoconfined superionic water (12 GPa) & 500~K & 500~ps & 432 & 1B, C; 2A, B, C; 3 & 1D (10 structures) \\
\hline
Nanoconfined water from Ref. 2 & 500~K (multiple temperatures) & 1~ns & 432 & 3B & \\
\hline
Nanoconfined superionic water (12 GPa) & 425~K, 450~K, 475~K, 500~K, 525~K, 550~K & 500~ps & 432 & S7 & \\
\hline
Bulk superionic water & 2500~K & 1~ns & 384 & 1A, C & 1D (10 structures) \\
\hline
Ice VII (50000~GPa) & 500~K & 2~ns & 864 & 3C & \\
\hline
Pressurised water (500~GPa) & 500~K & 2~ns & 864 & 3D & \\
\hline
Simulations of pressurised water and Ice VII in the NPT ensemble & 500~K & 500~ps & 864 & 3B & \\
\hline
Superionic AgI from ref 4 & 480~K & 1~ns & 2048 & S5, S6 & \\
\hline
\end{tabular}
\caption{Summary of simulations and corresponding figures which are plotted from them.}
\end{table}
\FloatBarrier
\section*{S3: Structure, bonding, and chain like diffusion in superionic silver iodide}

In the main paper in Fig.~1, we present structure and bonding calculations for bulk and nanoconfined superionic water and note the similarities and differences between the two systems. In Fig.~S6, we present the same information for superionic silver iodide. In general, we can observe similarities with the structure of BCC-ordered bulk superionic water. The snapshot in A and the form of the RDF in B are exceptionally similar, but a change in distances in the RDFs arises due to the far greater lattice constant of AgI.

We should, however, note that the form of the ICOBI plot is somewhat different from that of bulk superionic water. Though we see a similar distribution, it is centred at a similar level of ionic interactions. This difference can be attributed to two factors: firstly, the greater electronegativity gap between silver and iodine will lead to greater ionicity in bonding\,\cite{M_ller_2021}; and secondly, unlike hydrogen, silver is capable of having a far higher coordination number than two (the optimal value for a proton), leading to a greater number of weaker covalent interactions being formed with neighboring iodide ions\,\cite{Kreuer_1996}.

\begin{figure}[h]
    \centering
    \includegraphics[width=0.4\textwidth]{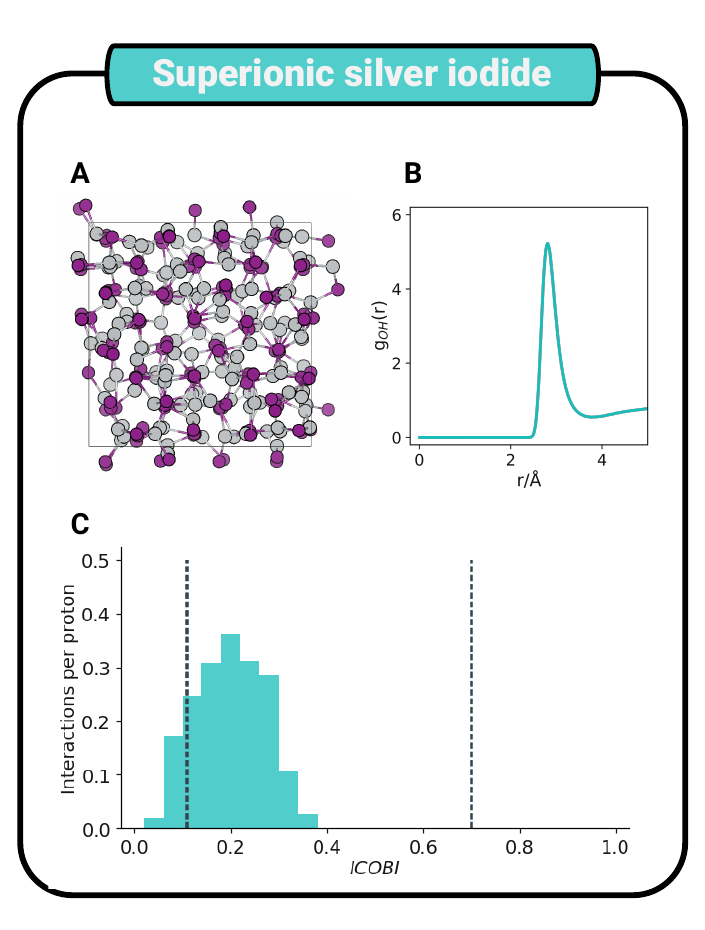}
    \caption{A companion figure to Fig. 1 in the main paper for superionic silver iodide. It comprises, (A) the structure of the phase, (B) the $g_{\mathrm{AgI}}(r)$ 480\,K, and (C) a plot of the ICOBI index for individual Ag–I for a snapshot taken at 480 K.}
\end{figure}

\begin{figure}[h]
    \centering
    \includegraphics[width=0.4\textwidth]{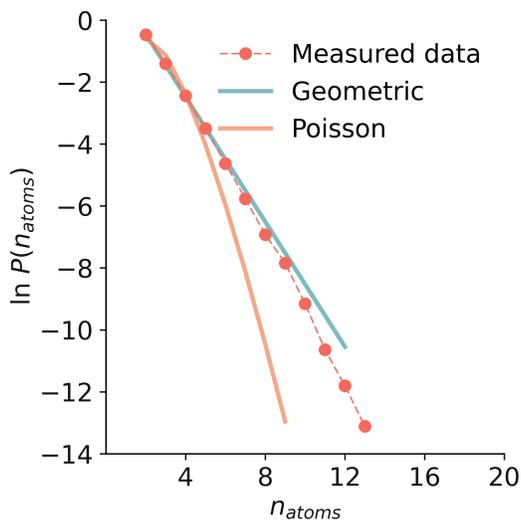}
    \caption{Plot of the log probability of chain lengths for silver iodide at 480~K. Geometric and Poisson fits to the data are shown, with the greater observed agreement with the geometric law indicating a chain-like diffusion mechanism. The parameters in this plot are taken as $\delta = 1.75$~\AA{} and $\Delta t = 1$~ps.}
\end{figure}

In addition to similarities in structure with bulk superionic water, we also observe chain-like diffusion in superionic silver iodide. In Fig.~S7, we investigate AgI using the same analysis as for the two superionic water-based systems in Fig.~2 of the main text, with $\delta = 1.75$~\AA{} and $\Delta t = 1$~ps. As with both forms of superionic water studied in the main paper, we observe better agreement with the geometric model. This is indicative of a chain-like diffusion mechanism.

\FloatBarrier
\section*{S4: Calculation of pKw from equilibrium simulations of nanoconfined superionic water}

\begin{figure}[h]
    \centering
    \includegraphics[width=0.4\textwidth]{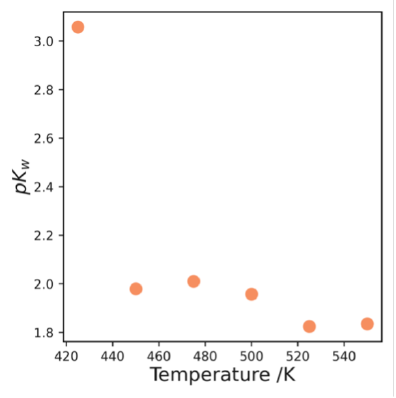}
    \caption{ Plots of $pK_\mathrm{w}$ calculated with temperature for nanoconfined superionic water, based on defect populations obtained from Voronoi tessellation.}
\end{figure}

The calculation of $pK_w$ from molecular simulations usually requires the employment of some form of advanced sampling technique. In nanoconfined superionic water, the level of ionisation is so high that the $pK_w$ can be calculated directly from the concentration of defect ions, using the Voronoi tessellation method for identifying defects. If we assume that the activity of pure nanoconfined water is

\begin{equation}
 p K_w=-\log _{10}\left(\left[\mathrm{OH}^{-}\right]\left[\mathrm{H}_3 \mathrm{O}^{+}\right]\right) .   
\end{equation}
In Fig.~S8 we report these results for nanoconfined superionic water from 420~K to 550~K. We observe values ranging from just above 3 to 1.8. These values are vastly different from the values previously obtained for nanoconfined water at atmospheric pressure using advanced sampling techniques\,\cite{Di_Pino_2023,Mu_oz_Santiburcio_2021}.

\FloatBarrier

\section*{S5: The effect of the inclusion of nuclear quantum effects on the structure of nanoconfined superionic water}

In this paper, we have focused on the qualitative description of the two types of superionic. While these two phases have been shown to be qualitatively different, it may be reasonably asked whether there is a profound change in the structure of nanoconfined superionic water when nuclear quantum effects (NQEs) are introduced. To these ends, we use simulations produced by Ravindra for Ref.\,2and compare the radial distribution function and the distribution of water bond angle with and without the inclusion of nuclear quantum effects (with the position of protons in the NQE inclusive plot taken as the positon of centroids). We observe in these plots shown in Fig.\,S8 that the inclusion of nuclear quantum effects, does not change the distributions beyond the expected broadening of peaks in both distributions. As a quantum hydrogen nuclei would be expected to be more diffuse in the bulk superionic phase as well, we can conclude that the qualitative differences described in this paper are not affected by quantum descriptions of nuclear motion.

\begin{figure}[h]
    \centering
    \includegraphics[width=0.4\textwidth]{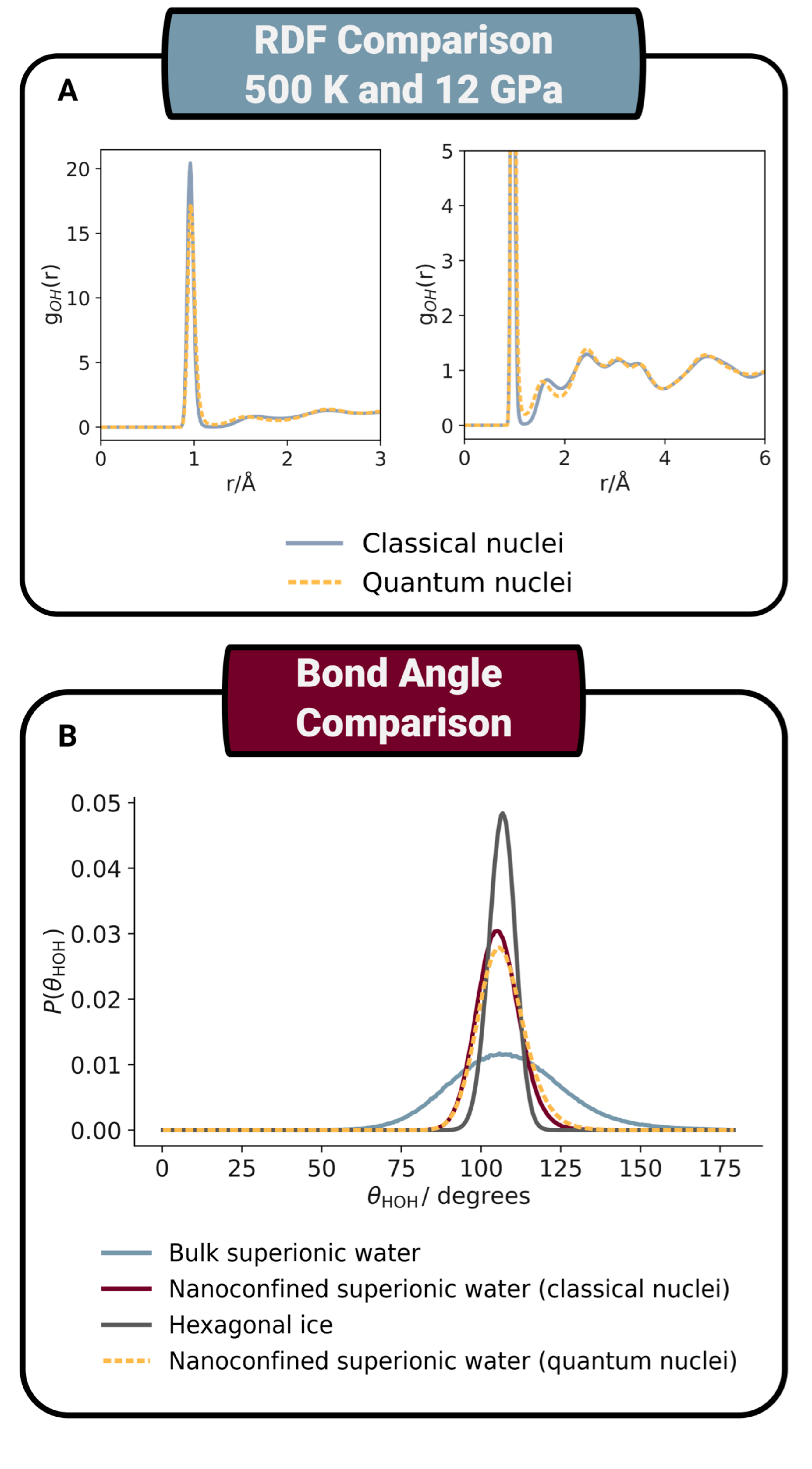}
    \caption{Plots exploring the effects of a quantum nuclear description on the structural descriptors presented for nanoconfined superionic water in the main paper. Panel (A) shows the effect of this description on the oxygen-hydrogen radial distribution function. To allow for the full appraisal of the the effect in structure a zoom of the full distribution is shown on the right hand side. Panel (B) reproduces the plot of the bond angle in the main paper with the values for quantum nuclei in nanoconfined superionic water added. In the NQE inclusive plots we plot the location of centroids as a proxy for the location of hydrogen nuclei.}
\end{figure}

\FloatBarrier
\section*{S6: Applying Voronoi tessellation to locate defects in nanoconfined superionic water}

The difference in the defect chemistry of nanoconfined and bulk superionic water becomes apparent if we try to allocate bulk defects as we do in the nanoconfined system, as shown in Fig.~S10. In the nanoconfined system, defects are allocated by Voronoi tessellation. When we do so, we identify that most oxygens have a coordination number of 2, with the small number of coordination numbers 1 and 3 representing the defects. 

If we do this on the bulk system, we obtain a broad distribution of coordination numbers, with 0 and 4 both widely observed. This is not demonstrative of a high number of defects, but rather that the nature of defects is fundamentally different, with bulk superionic water hydrogen defects being the conventional crystalline vacancies and interstitials. These are the conductive defects, instead of hydroxyl and hydronium ions.

\begin{figure}[h!]
    \centering
    \includegraphics[width=0.6\textwidth]{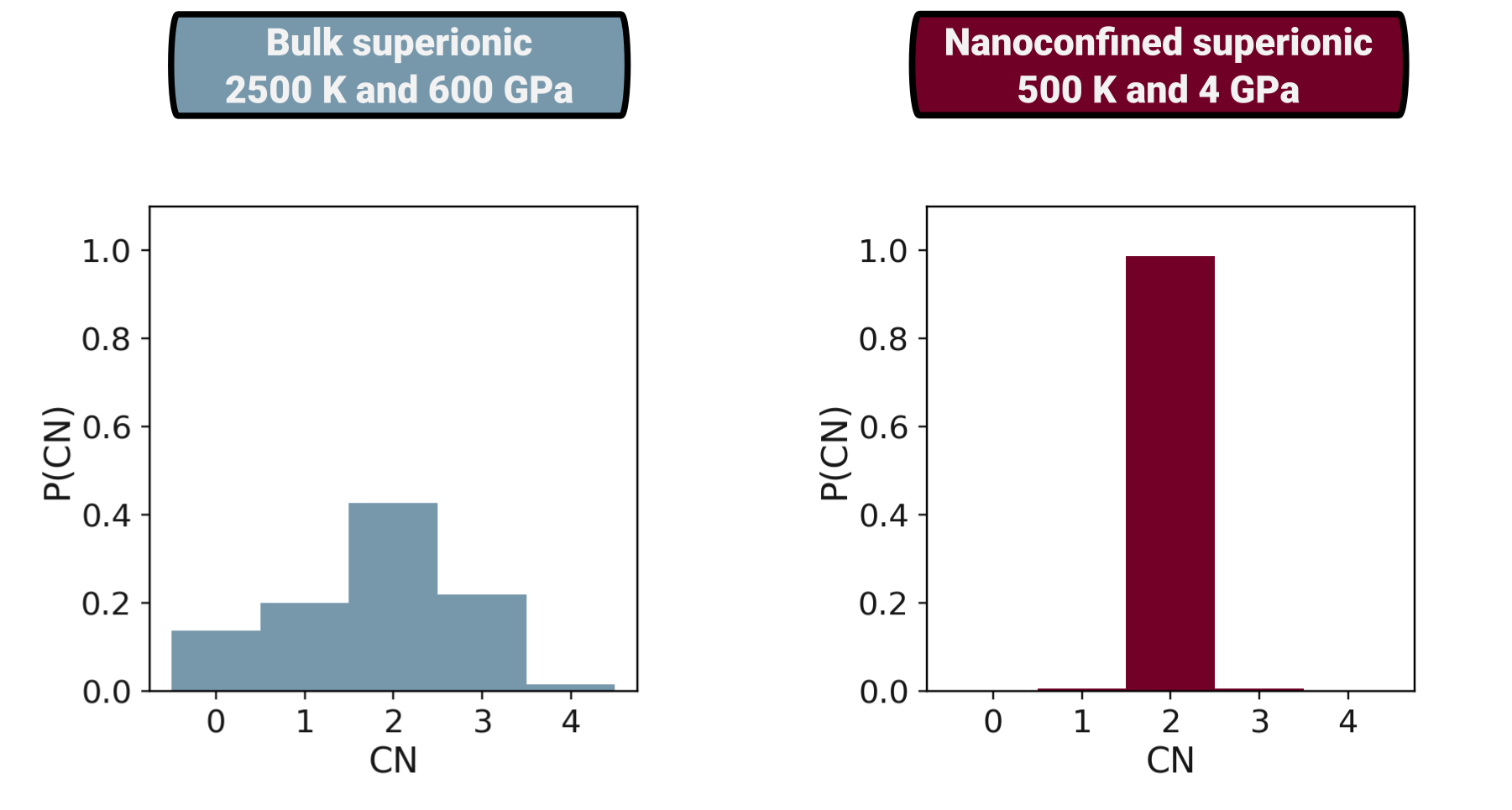}
    \caption{Plots of coordination numbers in the bulk and nanoconfined system obtained by Voronoi tessellation. Shown for bulk superionic water (2500~K, 600~GPa) and nanoconfined superionic water (600~K, 12~GPa).}
\end{figure}
\FloatBarrier

\section*{S7: Comparison of bulk superionic water neural network potential with hybrid DFT calculations}

In the main paper, we have compared and contrasted the structures and diffusive mechanisms of bulk superionic water and nanoconfined superionic water. In order to accurately simulate nanoconfined superionic water, Kapil et al.\ used a hybrid functional to fit the neural network potential\,\cite{Kapil_2022}. However, the bulk superionic potential used by Cheng\,\cite{Cheng_2021,Reinhardt_2022} and co-workers was fit from revPBE-D3 level DFT. 

\begin{figure}[h!]
    \centering
    \includegraphics[width=\textwidth]{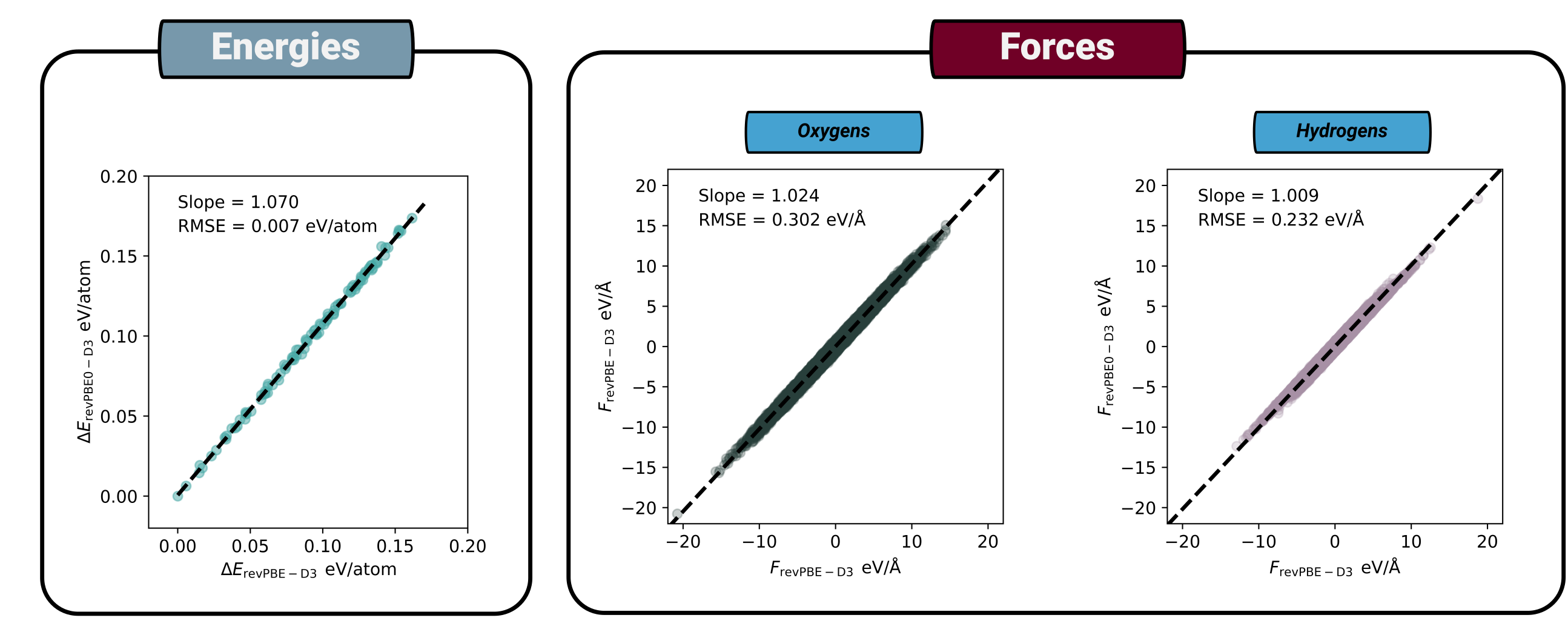}
    \caption{Plots of the energy and atomic forces of frames obtained with the bulk revPBE-D3 compared to revPBE-D3 calculations.}
\end{figure}

Though this potential was properly validated, it is worth considering if we would observe qualitatively different results if the potential had been fit with a hybrid DFT functional. To consider this, we have taken structures from a trajectory of bulk superionic water at 2500~K and 600~GPa. Structures were selected from the trajectory by ranking structures by energy and taking 100 structures at equal increments along this energy scale. 

As the model was fit using plane wave DFT calculation we perform the comparison in energies and forces using the Density functional theory calculations were performed using the CP2K code at the revPBE level used to fit the model and th revPBE0-D3 level of theory with the settings used to fit the model of Kapil \textit{et al.}\,\cite{Kapil_2022} used for both sets of calculations. We plot the resulting energies of frames against each other in Fig.~S11, where we see a good agreement in the energetic ordering of structures between functionals; though, of course, the RMSE errors are substantially greater than those obtained during a fitting process. 

Further, the slope of the line of best fit is close to one; this suggests the overall width of the configurational density of states relative to the thermal energy is similar for systems with interactions described by the two functionals. This suggests that the physics described by the neural network potential should be qualitatively similar to that we would obtain at a hybrid level of theory. 

We further observe similar agreement for forces acting on hydrogens and oxygens in these frames. This suggests that the potential energy surfaces described at these two levels of theory are similar enough in form and lend force to the qualitative conclusions we have drawn in this paper.
\FloatBarrier
\section*{S8: Sensitivity-based analysis of the parameters for calculating chains}

In this paper, we have considered the possibility of concerted diffusion mechanisms from the perspective of the chain-like diffusion descriptors previously used to describe diffusion in glasses\,\cite{Donati_1998} and solid electrolytes\,\cite{Morgan_2021}. Chains are identified on a pairwise basis, where one atom---in our case limited to the hydrogen atoms in the system---has moved away from its initial site and been replaced by another. This is defined mathematically for atoms with positions $\mathbf{r}_i$ and $\mathbf{r}_j$ at time $t$ as,

\begin{equation}
\min \left[\left\|r_{\mathrm{i}}(t+\Delta t)-r_j(t)\right\|,\left\|r_j(t+\Delta t)-r_j(t)\right\|\right]<\delta
\end{equation}

For a time window $\Delta t$ and a spatial cutoff $\delta$, the value of these variables needs to be set and has a quantitative effect on the length of chains. However, the qualitative results obtained have previously been suggested to have limited dependence on these parameters in Lennard-Jones glasses, provided that the spatial cutoff is smaller than the hard sphere radius. In crystalline solid electrolytes, this cutoff can be meaningfully set based on the periodicity of the framework lattice. Given the two-step diffusion mechanism of nanoconfined superionic water, the exact setting of the spatial cutoff is difficult, and definitions based on both the excluded volume of protons and water molecules can be argued for. To assess whether the qualitative conclusions in the main paper are robust to the choice of these values, we performed a sensitivity analysis. Regardless of the chosen values, we observe a geometric decay in probability in chain lengths for both nanoconfined and bulk superionic water, giving a strong degree of confidence in both cases that diffusion is chain-like. Plots are shown below for both varieties of superionic water for spatial cutoffs of $1\,\text{\AA}$, $1.5\,\text{\AA}$, and $2\,\text{\AA}$ for multiple time windows for each system.

\newpage
\subsection*{Nanoconfined superionic water}
\FloatBarrier
\begin{figure}[b!]
    \centering
    \includegraphics[width=0.8\textwidth]{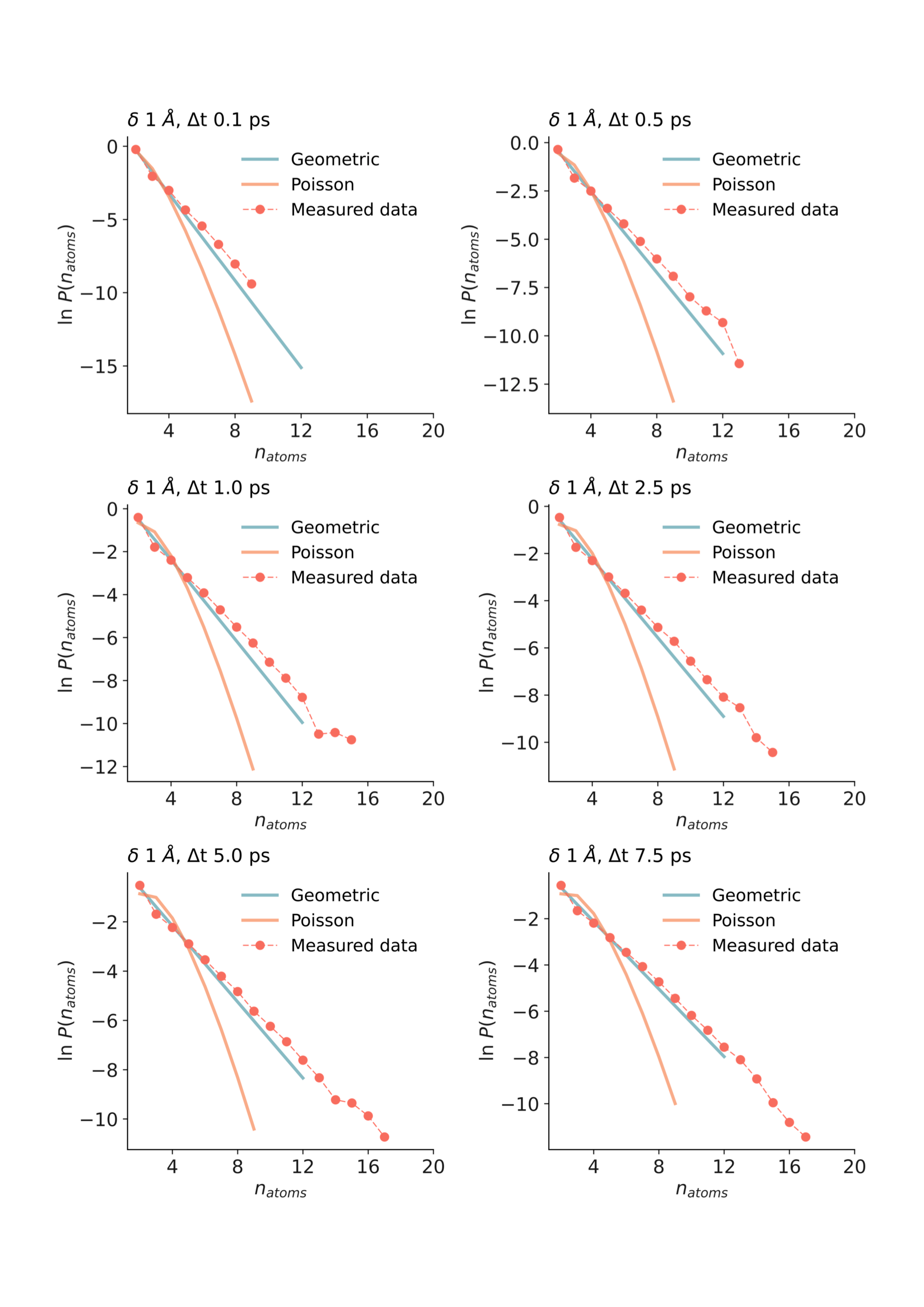}
    \caption{Plots of the log probability of the lengths of diffusive chains for nanoconfined superionic water with a spatial cutoff ($\delta$) of 1~\AA, and time windows of $\Delta t$ of: 0.1~ps, 0.5~ps, 1.0~ps, 2.5~ps, 5.0~ps, and 7.5~ps. Fits to Geometric and Poisson models are provided; these models are representative of chain-like and isolated hopping-based diffusion.}
    \label{fig:S12}
\end{figure}

\begin{figure}[h!]
    \centering
    \includegraphics[width=0.8\textwidth]{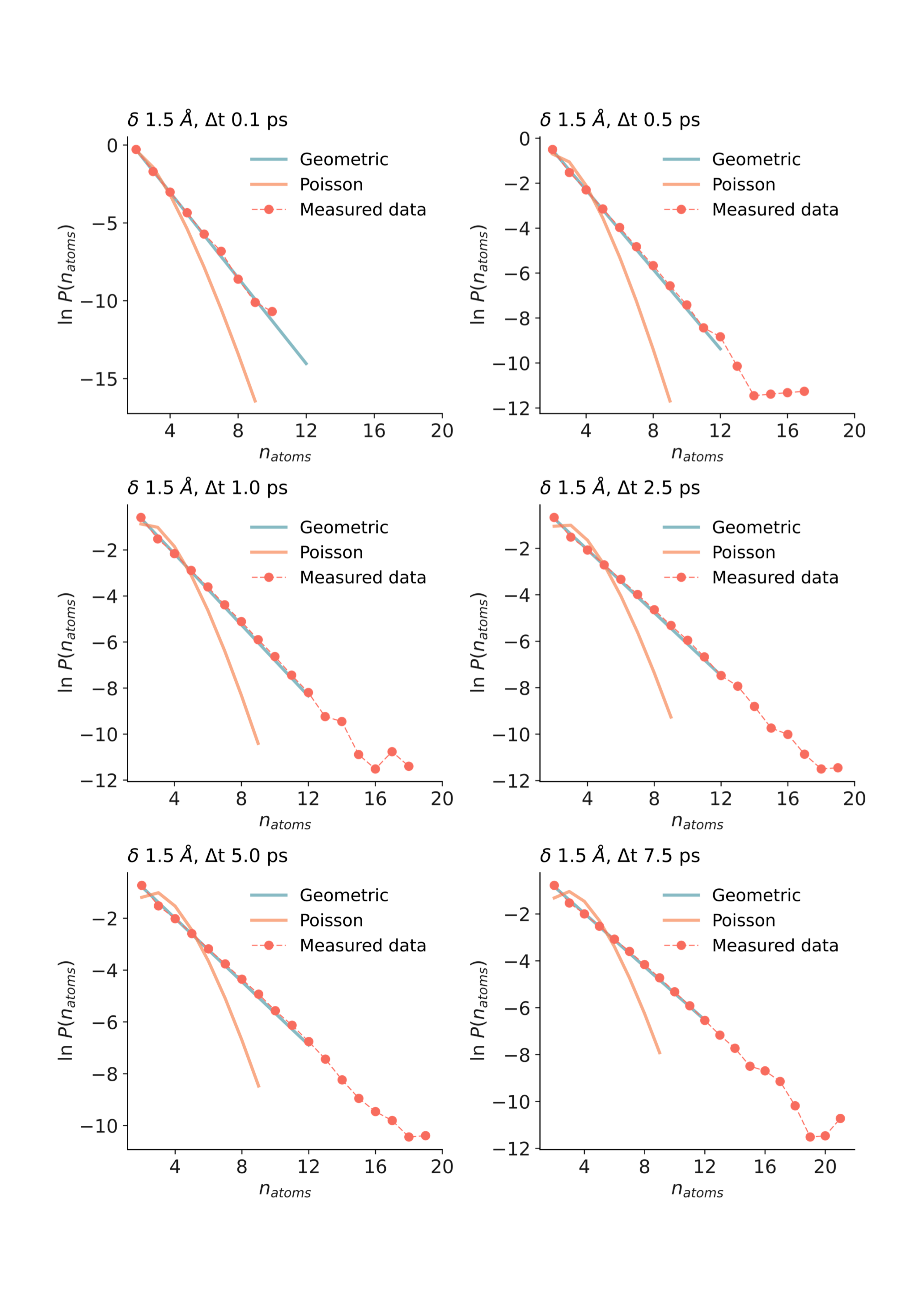}
    \caption{ Plots of the log probability of the lengths of diffusive chains for nanoconfined superionic water with a spatial cutoff ($\delta$) of 1.5~\AA, and time windows of $\Delta t$ of: 0.1~ps, 0.5~ps, 1.0~ps, 2.5~ps, 5.0~ps, and 7.5~ps. Fits to Geometric and Poisson models are provided; these models are representative of chain-like and isolated hopping-based diffusion.}
    \label{fig:S12}
\end{figure}

\begin{figure}[h!]
    \centering
    \includegraphics[width=0.8\textwidth]{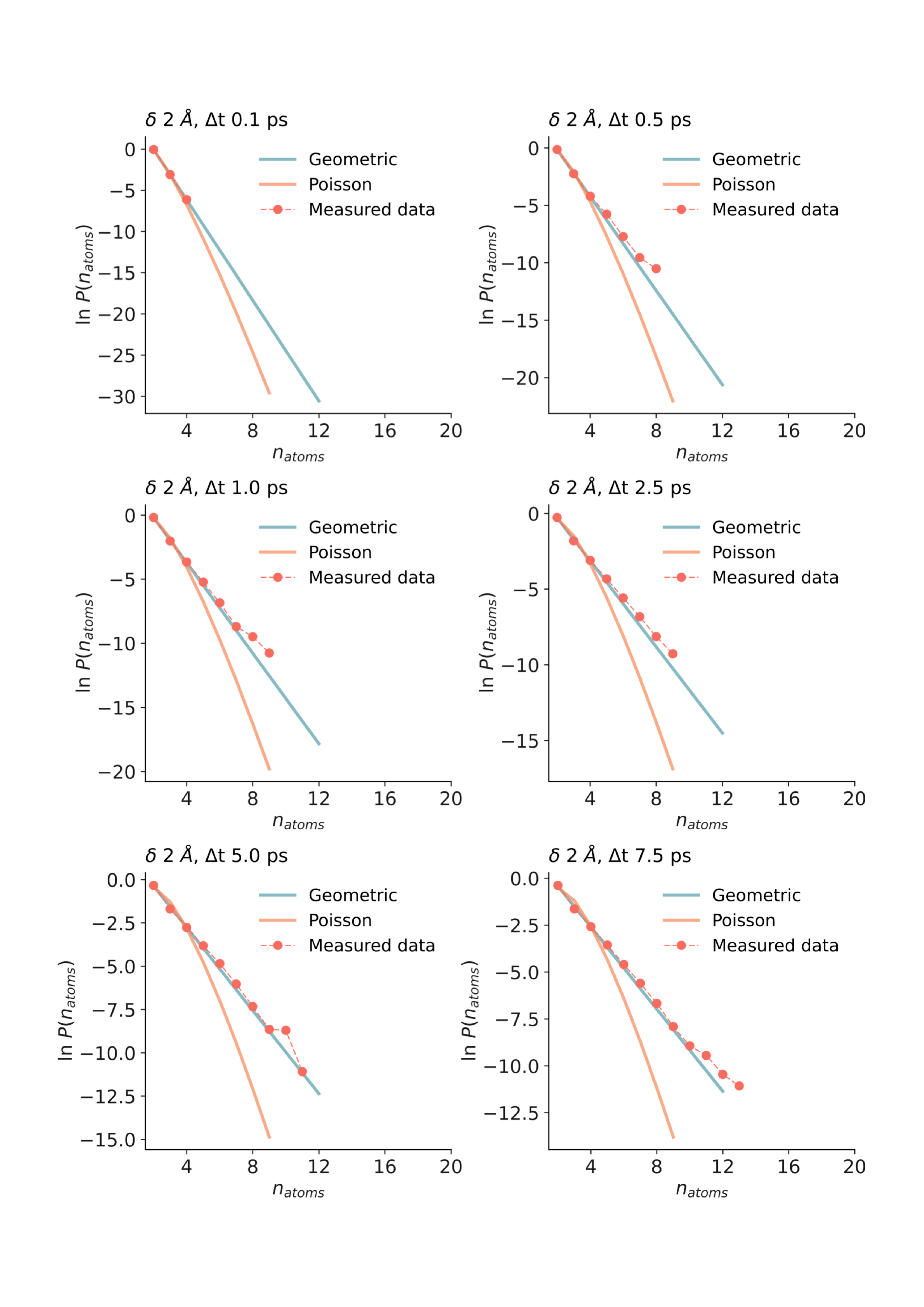}
    \caption{Plots of the log probability of the lengths of diffusive chains for nanoconfined superionic water with a spatial cutoff ($\delta$) of 2~\AA, and time windows of $\Delta t$ of: 0.1~ps, 0.5~ps, 1.0~ps, 2.5~ps, 5.0~ps, and 7.5~ps. Fits to Geometric and Poisson models are provided; these models are representative of chain-like and isolated hopping-based diffusion.}
    \label{fig:S13}
\end{figure}
\newpage
\newpage
\FloatBarrier
\subsection*{Bulk superionic water}

\begin{figure}[h!]
    \centering
    \includegraphics[width=0.8\textwidth]{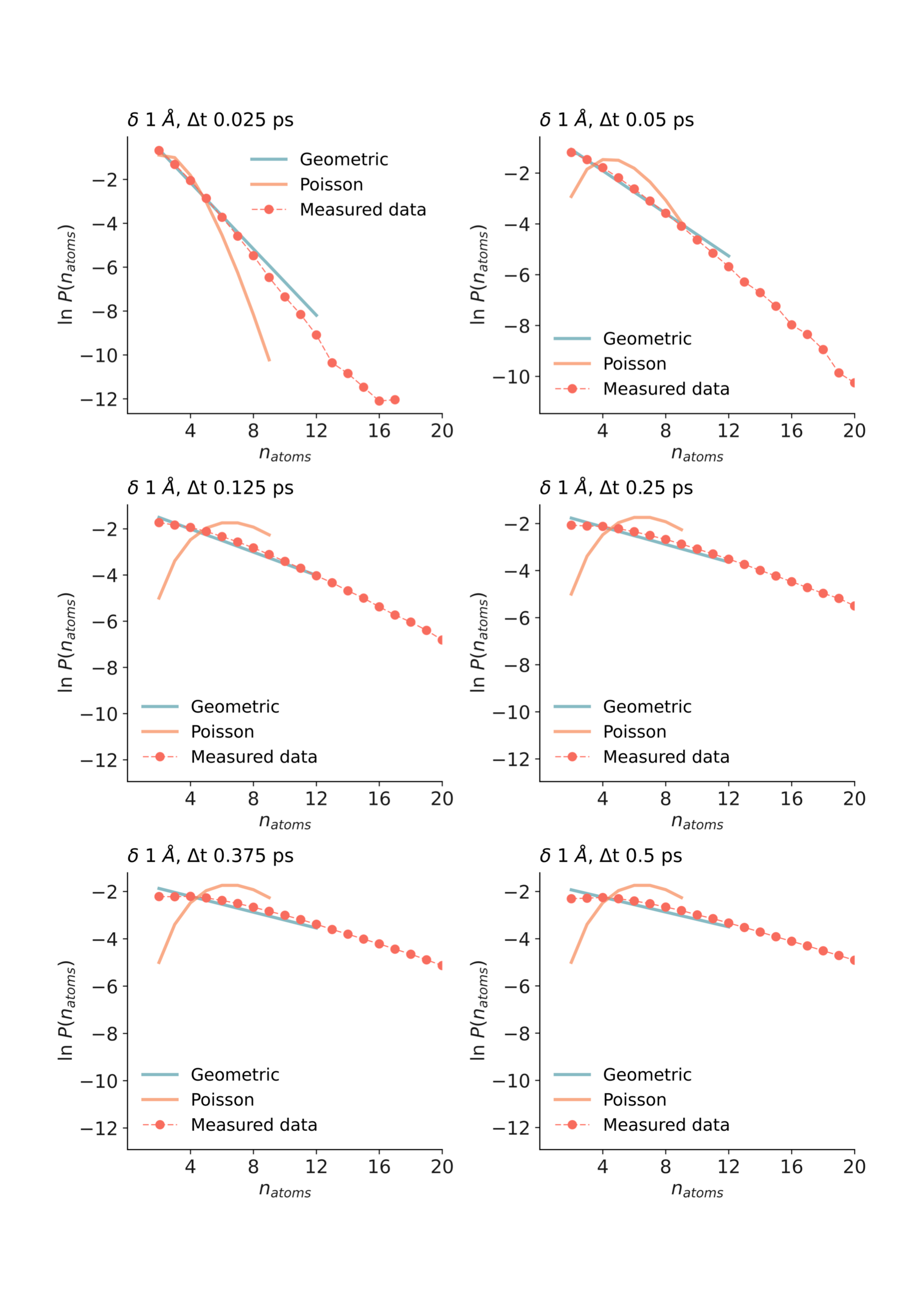}
    \caption{Plots of the log probability of the lengths of diffusive chains for bulk superionic water with a spatial cutoff ($\delta$) of 1~\AA, and time windows of $\Delta t$ of: 0.025~ps, 0.05~ps, 0.125~ps, 0.25~ps, 0.375~ps, and 0.5~ps. Fits to Geometric and Poisson models are provided; these models are representative of chain-like and isolated hopping-based diffusion.}
    \label{fig:S14}
\end{figure}
\newpage
\begin{figure}[h!]
    \centering
    \includegraphics[width=0.8\textwidth]{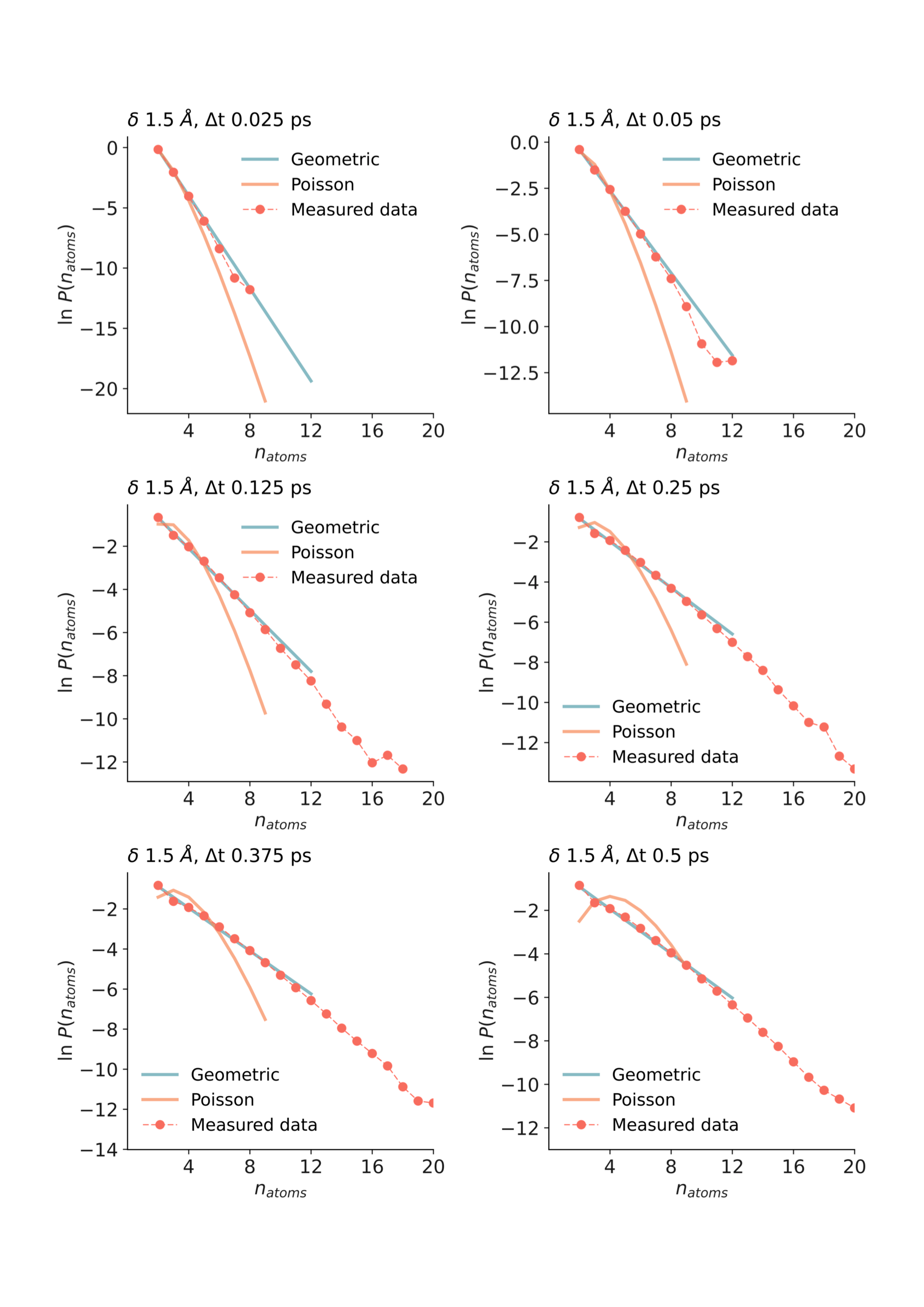}
    \caption{Plots of the log probability of the lengths of diffusive chains for bulk superionic water with a spatial cutoff ($\delta$) of 1.5~\AA, and time windows of $\Delta t$ of: 0.025~ps, 0.05~ps, 0.125~ps, 0.25~ps, 0.375~ps, and 0.5~ps. Fits to Geometric and Poisson models are provided; these models are representative of chain-like and isolated hopping-based diffusion.}
    \label{fig:S15}
\end{figure}
\newpage
\begin{figure}[h!]
    \centering
    \includegraphics[width=0.8\textwidth]{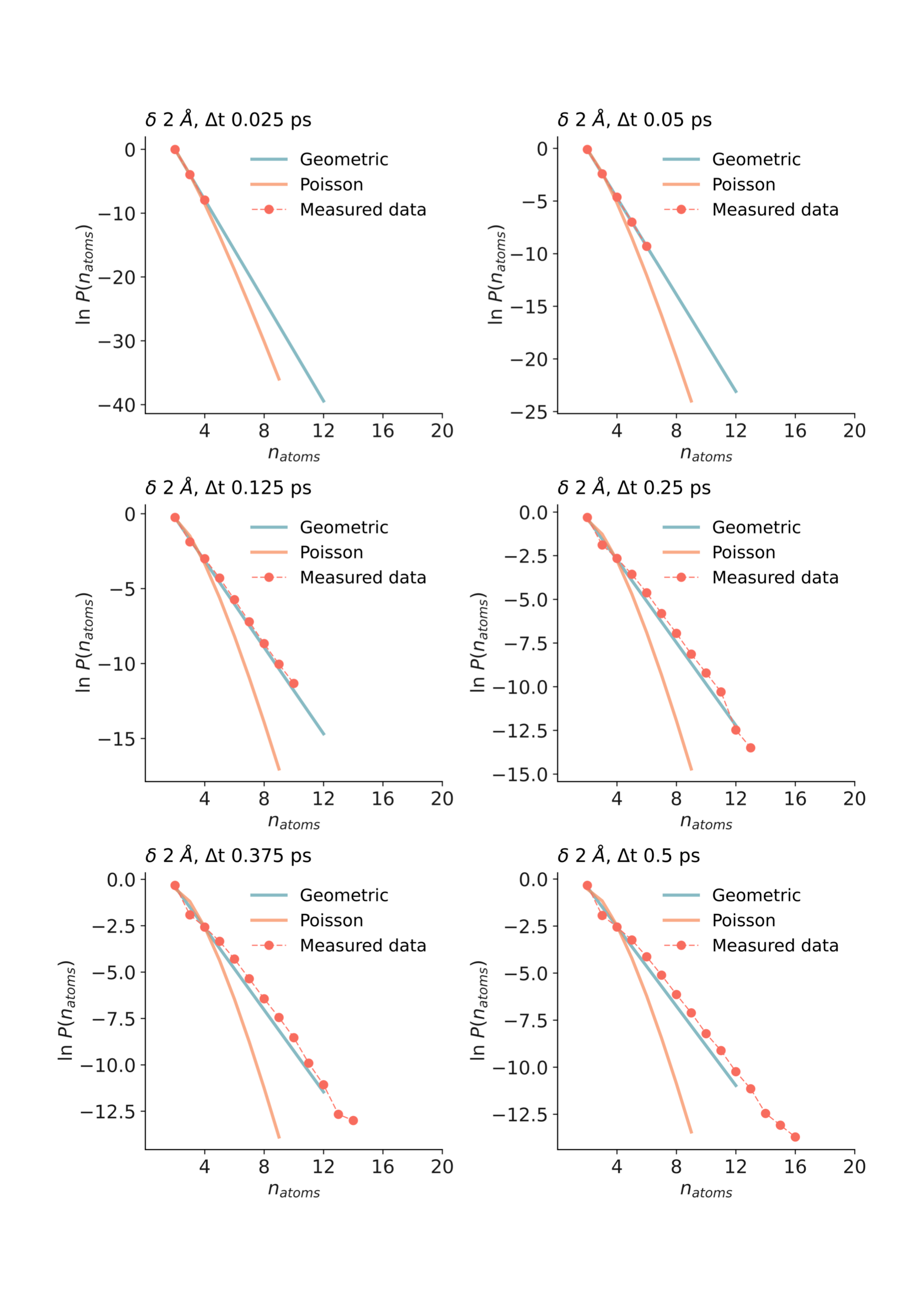}
    \caption{Plots of the log probability of the lengths of diffusive chains for bulk superionic water with a spatial cutoff ($\delta$) of 2~\AA, and time windows of $\Delta t$ of: 0.025~ps, 0.05~ps, 0.125~ps, 0.25~ps, 0.375~ps, and 0.5~ps. Fits to Geometric and Poisson models are provided; these models are representative of chain-like and isolated hopping-based diffusion.}
    \label{fig:S16}
\end{figure}

\FloatBarrier

\section*{References}
\begingroup
\renewcommand{\section}[2]{} 

\endgroup


\begin{thebibliography}{68}%
\makeatletter
\providecommand \@ifxundefined [1]{%
 \@ifx{#1\undefined}
}%
\providecommand \@ifnum [1]{%
 \ifnum #1\expandafter \@firstoftwo
 \else \expandafter \@secondoftwo
 \fi
}%
\providecommand \@ifx [1]{%
 \ifx #1\expandafter \@firstoftwo
 \else \expandafter \@secondoftwo
 \fi
}%
\providecommand \natexlab [1]{#1}%
\providecommand \enquote  [1]{``#1''}%
\providecommand \bibnamefont  [1]{#1}%
\providecommand \bibfnamefont [1]{#1}%
\providecommand \citenamefont [1]{#1}%
\providecommand \href@noop [0]{\@secondoftwo}%
\providecommand \href [0]{\begingroup \@sanitize@url \@href}%
\providecommand \@href[1]{\@@startlink{#1}\@@href}%
\providecommand \@@href[1]{\endgroup#1\@@endlink}%
\providecommand \@sanitize@url [0]{\catcode `\\12\catcode `\$12\catcode `\&12\catcode `\#12\catcode `\^12\catcode `\_12\catcode `\%12\relax}%
\providecommand \@@startlink[1]{}%
\providecommand \@@endlink[0]{}%
\providecommand \url  [0]{\begingroup\@sanitize@url \@url }%
\providecommand \@url [1]{\endgroup\@href {#1}{\urlprefix }}%
\providecommand \urlprefix  [0]{URL }%
\providecommand \Eprint [0]{\href }%
\providecommand \doibase [0]{https://doi.org/}%
\providecommand \selectlanguage [0]{\@gobble}%
\providecommand \bibinfo  [0]{\@secondoftwo}%
\providecommand \bibfield  [0]{\@secondoftwo}%
\providecommand \translation [1]{[#1]}%
\providecommand \BibitemOpen [0]{}%
\providecommand \bibitemStop [0]{}%
\providecommand \bibitemNoStop [0]{.\EOS\space}%
\providecommand \EOS [0]{\spacefactor3000\relax}%
\providecommand \BibitemShut  [1]{\csname bibitem#1\endcsname}%
\let\auto@bib@innerbib\@empty
\bibitem [{\citenamefont {Funke}(2013)}]{Funke2013}%
  \BibitemOpen
  \bibfield  {author} {\bibinfo {author} {\bibfnamefont {K.}~\bibnamefont {Funke}},\ }\href {https://doi.org/10.1088/1468-6996/14/4/043502} {\bibfield  {journal} {\bibinfo  {journal} {Science and Technology of Advanced Materials}\ }\textbf {\bibinfo {volume} {14}},\ \bibinfo {pages} {043502} (\bibinfo {year} {2013})}\BibitemShut {NoStop}%
\bibitem [{\citenamefont {Faraday}(1843)}]{Faraday_1843}%
  \BibitemOpen
  \bibfield  {author} {\bibinfo {author} {\bibfnamefont {M.}~\bibnamefont {Faraday}},\ }\href {https://doi.org/10.1098/rspl.1837.0015} {\bibfield  {journal} {\bibinfo  {journal} {Abstracts of the Papers Printed in the Philosophical Transactions of the Royal Society of London}\ }\textbf {\bibinfo {volume} {4}},\ \bibinfo {pages} {49} (\bibinfo {year} {1843})}\BibitemShut {NoStop}%
\bibitem [{\citenamefont {Rice}\ and\ \citenamefont {Roth}(1972)}]{RICE1972294}%
  \BibitemOpen
  \bibfield  {author} {\bibinfo {author} {\bibfnamefont {M.}~\bibnamefont {Rice}}\ and\ \bibinfo {author} {\bibfnamefont {W.}~\bibnamefont {Roth}},\ }\href {https://doi.org/https://doi.org/10.1016/0022-4596(72)90121-1} {\bibfield  {journal} {\bibinfo  {journal} {Journal of Solid State Chemistry}\ }\textbf {\bibinfo {volume} {4}},\ \bibinfo {pages} {294} (\bibinfo {year} {1972})}\BibitemShut {NoStop}%
\bibitem [{\citenamefont {Thangadurai}\ \emph {et~al.}(2003)\citenamefont {Thangadurai}, \citenamefont {Kaack},\ and\ \citenamefont {Weppner}}]{Thangadurai_2003}%
  \BibitemOpen
  \bibfield  {author} {\bibinfo {author} {\bibfnamefont {V.}~\bibnamefont {Thangadurai}}, \bibinfo {author} {\bibfnamefont {H.}~\bibnamefont {Kaack}},\ and\ \bibinfo {author} {\bibfnamefont {W.~J.~F.}\ \bibnamefont {Weppner}},\ }\href {https://doi.org/10.1111/j.1151-2916.2003.tb03318.x} {\bibfield  {journal} {\bibinfo  {journal} {Journal of the American Ceramic Society}\ }\textbf {\bibinfo {volume} {86}},\ \bibinfo {pages} {437–440} (\bibinfo {year} {2003})}\BibitemShut {NoStop}%
\bibitem [{\citenamefont {Murugan}\ \emph {et~al.}(2007)\citenamefont {Murugan}, \citenamefont {Thangadurai},\ and\ \citenamefont {Weppner}}]{Murugan_2007}%
  \BibitemOpen
  \bibfield  {author} {\bibinfo {author} {\bibfnamefont {R.}~\bibnamefont {Murugan}}, \bibinfo {author} {\bibfnamefont {V.}~\bibnamefont {Thangadurai}},\ and\ \bibinfo {author} {\bibfnamefont {W.}~\bibnamefont {Weppner}},\ }\href {https://doi.org/10.1002/anie.200701144} {\bibfield  {journal} {\bibinfo  {journal} {Angewandte Chemie International Edition}\ }\textbf {\bibinfo {volume} {46}},\ \bibinfo {pages} {7778–7781} (\bibinfo {year} {2007})}\BibitemShut {NoStop}%
\bibitem [{\citenamefont {Kamaya}\ \emph {et~al.}(2011)\citenamefont {Kamaya}, \citenamefont {Homma}, \citenamefont {Yamakawa}, \citenamefont {Hirayama}, \citenamefont {Kanno}, \citenamefont {Yonemura}, \citenamefont {Kamiyama}, \citenamefont {Kato}, \citenamefont {Hama}, \citenamefont {Kawamoto},\ and\ \citenamefont {Mitsui}}]{Kamaya_2011}%
  \BibitemOpen
  \bibfield  {author} {\bibinfo {author} {\bibfnamefont {N.}~\bibnamefont {Kamaya}}, \bibinfo {author} {\bibfnamefont {K.}~\bibnamefont {Homma}}, \bibinfo {author} {\bibfnamefont {Y.}~\bibnamefont {Yamakawa}}, \bibinfo {author} {\bibfnamefont {M.}~\bibnamefont {Hirayama}}, \bibinfo {author} {\bibfnamefont {R.}~\bibnamefont {Kanno}}, \bibinfo {author} {\bibfnamefont {M.}~\bibnamefont {Yonemura}}, \bibinfo {author} {\bibfnamefont {T.}~\bibnamefont {Kamiyama}}, \bibinfo {author} {\bibfnamefont {Y.}~\bibnamefont {Kato}}, \bibinfo {author} {\bibfnamefont {S.}~\bibnamefont {Hama}}, \bibinfo {author} {\bibfnamefont {K.}~\bibnamefont {Kawamoto}},\ and\ \bibinfo {author} {\bibfnamefont {A.}~\bibnamefont {Mitsui}},\ }\href {https://doi.org/10.1038/nmat3066} {\bibfield  {journal} {\bibinfo  {journal} {Nature Materials}\ }\textbf {\bibinfo {volume} {10}},\ \bibinfo {pages} {682–686} (\bibinfo {year} {2011})}\BibitemShut {NoStop}%
\bibitem [{\citenamefont {Paren}\ \emph {et~al.}(2022)\citenamefont {Paren}, \citenamefont {Nguyen}, \citenamefont {Ballance}, \citenamefont {Hallinan}, \citenamefont {Kennemur},\ and\ \citenamefont {Winey}}]{Paren_2022}%
  \BibitemOpen
  \bibfield  {author} {\bibinfo {author} {\bibfnamefont {B.~A.}\ \bibnamefont {Paren}}, \bibinfo {author} {\bibfnamefont {N.}~\bibnamefont {Nguyen}}, \bibinfo {author} {\bibfnamefont {V.}~\bibnamefont {Ballance}}, \bibinfo {author} {\bibfnamefont {D.~T.}\ \bibnamefont {Hallinan}}, \bibinfo {author} {\bibfnamefont {J.~G.}\ \bibnamefont {Kennemur}},\ and\ \bibinfo {author} {\bibfnamefont {K.~I.}\ \bibnamefont {Winey}},\ }\href {https://doi.org/10.1021/acs.macromol.2c00459} {\bibfield  {journal} {\bibinfo  {journal} {Macromolecules}\ }\textbf {\bibinfo {volume} {55}},\ \bibinfo {pages} {4692–4702} (\bibinfo {year} {2022})}\BibitemShut {NoStop}%
\bibitem [{\citenamefont {Wang}\ and\ \citenamefont {Sokolov}(2015)}]{Wang_2015}%
  \BibitemOpen
  \bibfield  {author} {\bibinfo {author} {\bibfnamefont {Y.}~\bibnamefont {Wang}}\ and\ \bibinfo {author} {\bibfnamefont {A.~P.}\ \bibnamefont {Sokolov}},\ }\href {https://doi.org/10.1016/j.coche.2014.09.002} {\bibfield  {journal} {\bibinfo  {journal} {Current Opinion in Chemical Engineering}\ }\textbf {\bibinfo {volume} {7}},\ \bibinfo {pages} {113–119} (\bibinfo {year} {2015})}\BibitemShut {NoStop}%
\bibitem [{\citenamefont {Millot}\ \emph {et~al.}(2018)\citenamefont {Millot}, \citenamefont {Hamel}, \citenamefont {Rygg}, \citenamefont {Celliers}, \citenamefont {Collins}, \citenamefont {Coppari}, \citenamefont {Fratanduono}, \citenamefont {Jeanloz}, \citenamefont {Swift},\ and\ \citenamefont {Eggert}}]{Millot_2018}%
  \BibitemOpen
  \bibfield  {author} {\bibinfo {author} {\bibfnamefont {M.}~\bibnamefont {Millot}}, \bibinfo {author} {\bibfnamefont {S.}~\bibnamefont {Hamel}}, \bibinfo {author} {\bibfnamefont {J.~R.}\ \bibnamefont {Rygg}}, \bibinfo {author} {\bibfnamefont {P.~M.}\ \bibnamefont {Celliers}}, \bibinfo {author} {\bibfnamefont {G.~W.}\ \bibnamefont {Collins}}, \bibinfo {author} {\bibfnamefont {F.}~\bibnamefont {Coppari}}, \bibinfo {author} {\bibfnamefont {D.~E.}\ \bibnamefont {Fratanduono}}, \bibinfo {author} {\bibfnamefont {R.}~\bibnamefont {Jeanloz}}, \bibinfo {author} {\bibfnamefont {D.~C.}\ \bibnamefont {Swift}},\ and\ \bibinfo {author} {\bibfnamefont {J.~H.}\ \bibnamefont {Eggert}},\ }\href {https://doi.org/10.1038/s41567-017-0017-4} {\bibfield  {journal} {\bibinfo  {journal} {Nature Physics}\ }\textbf {\bibinfo {volume} {14}},\ \bibinfo {pages} {297–302} (\bibinfo {year} {2018})}\BibitemShut {NoStop}%
\bibitem [{\citenamefont {Demontis}\ \emph {et~al.}(1988)\citenamefont {Demontis}, \citenamefont {LeSar},\ and\ \citenamefont {Klein}}]{Demontis_1988}%
  \BibitemOpen
  \bibfield  {author} {\bibinfo {author} {\bibfnamefont {P.}~\bibnamefont {Demontis}}, \bibinfo {author} {\bibfnamefont {R.}~\bibnamefont {LeSar}},\ and\ \bibinfo {author} {\bibfnamefont {M.~L.}\ \bibnamefont {Klein}},\ }\href {https://doi.org/10.1103/physrevlett.60.2284} {\bibfield  {journal} {\bibinfo  {journal} {Physical Review Letters}\ }\textbf {\bibinfo {volume} {60}},\ \bibinfo {pages} {2284–2287} (\bibinfo {year} {1988})}\BibitemShut {NoStop}%
\bibitem [{\citenamefont {Prakapenka}\ \emph {et~al.}(2021)\citenamefont {Prakapenka}, \citenamefont {Holtgrewe}, \citenamefont {Lobanov},\ and\ \citenamefont {Goncharov}}]{Prakapenka_2021}%
  \BibitemOpen
  \bibfield  {author} {\bibinfo {author} {\bibfnamefont {V.~B.}\ \bibnamefont {Prakapenka}}, \bibinfo {author} {\bibfnamefont {N.}~\bibnamefont {Holtgrewe}}, \bibinfo {author} {\bibfnamefont {S.~S.}\ \bibnamefont {Lobanov}},\ and\ \bibinfo {author} {\bibfnamefont {A.~F.}\ \bibnamefont {Goncharov}},\ }\href {https://doi.org/10.1038/s41567-021-01351-8} {\bibfield  {journal} {\bibinfo  {journal} {Nature Physics}\ }\textbf {\bibinfo {volume} {17}},\ \bibinfo {pages} {1233–1238} (\bibinfo {year} {2021})}\BibitemShut {NoStop}%
\bibitem [{\citenamefont {Cheng}\ \emph {et~al.}(2021)\citenamefont {Cheng}, \citenamefont {Bethkenhagen}, \citenamefont {Pickard},\ and\ \citenamefont {Hamel}}]{Cheng_2021}%
  \BibitemOpen
  \bibfield  {author} {\bibinfo {author} {\bibfnamefont {B.}~\bibnamefont {Cheng}}, \bibinfo {author} {\bibfnamefont {M.}~\bibnamefont {Bethkenhagen}}, \bibinfo {author} {\bibfnamefont {C.~J.}\ \bibnamefont {Pickard}},\ and\ \bibinfo {author} {\bibfnamefont {S.}~\bibnamefont {Hamel}},\ }\href {https://doi.org/10.1038/s41567-021-01334-9} {\bibfield  {journal} {\bibinfo  {journal} {Nature Physics}\ }\textbf {\bibinfo {volume} {17}},\ \bibinfo {pages} {1228–1232} (\bibinfo {year} {2021})}\BibitemShut {NoStop}%
\bibitem [{\citenamefont {Goldman}\ \emph {et~al.}(2005)\citenamefont {Goldman}, \citenamefont {Fried}, \citenamefont {Kuo},\ and\ \citenamefont {Mundy}}]{Goldman_2005}%
  \BibitemOpen
  \bibfield  {author} {\bibinfo {author} {\bibfnamefont {N.}~\bibnamefont {Goldman}}, \bibinfo {author} {\bibfnamefont {L.~E.}\ \bibnamefont {Fried}}, \bibinfo {author} {\bibfnamefont {I.-F.~W.}\ \bibnamefont {Kuo}},\ and\ \bibinfo {author} {\bibfnamefont {C.~J.}\ \bibnamefont {Mundy}},\ }\href {https://doi.org/10.1103/PhysRevLett.94.217801} {\bibfield  {journal} {\bibinfo  {journal} {Physical Review Letters}\ }\textbf {\bibinfo {volume} {94}},\ \bibinfo {pages} {217801} (\bibinfo {year} {2005})}\BibitemShut {NoStop}%
\bibitem [{\citenamefont {Matusalem}\ \emph {et~al.}(2022)\citenamefont {Matusalem}, \citenamefont {Santos~Rego},\ and\ \citenamefont {de~Koning}}]{Matusalem_2022}%
  \BibitemOpen
  \bibfield  {author} {\bibinfo {author} {\bibfnamefont {F.}~\bibnamefont {Matusalem}}, \bibinfo {author} {\bibfnamefont {J.}~\bibnamefont {Santos~Rego}},\ and\ \bibinfo {author} {\bibfnamefont {M.}~\bibnamefont {de~Koning}},\ }\href {https://doi.org/10.1073/pnas.2203397119} {\bibfield  {journal} {\bibinfo  {journal} {Proceedings of the National Academy of Sciences}\ }\textbf {\bibinfo {volume} {119}},\ \bibinfo {pages} {e2203397119} (\bibinfo {year} {2022})}\BibitemShut {NoStop}%
\bibitem [{\citenamefont {Kapil}\ \emph {et~al.}(2022)\citenamefont {Kapil}, \citenamefont {Schran}, \citenamefont {Zen}, \citenamefont {Chen}, \citenamefont {Pickard},\ and\ \citenamefont {Michaelides}}]{Kapil_2022}%
  \BibitemOpen
  \bibfield  {author} {\bibinfo {author} {\bibfnamefont {V.}~\bibnamefont {Kapil}}, \bibinfo {author} {\bibfnamefont {C.}~\bibnamefont {Schran}}, \bibinfo {author} {\bibfnamefont {A.}~\bibnamefont {Zen}}, \bibinfo {author} {\bibfnamefont {J.}~\bibnamefont {Chen}}, \bibinfo {author} {\bibfnamefont {C.~J.}\ \bibnamefont {Pickard}},\ and\ \bibinfo {author} {\bibfnamefont {A.}~\bibnamefont {Michaelides}},\ }\href {https://doi.org/10.1038/s41586-022-05036-x} {\bibfield  {journal} {\bibinfo  {journal} {Nature}\ }\textbf {\bibinfo {volume} {609}},\ \bibinfo {pages} {512–516} (\bibinfo {year} {2022})}\BibitemShut {NoStop}%
\bibitem [{\citenamefont {Algara-Siller}\ \emph {et~al.}(2015)\citenamefont {Algara-Siller}, \citenamefont {Lehtinen}, \citenamefont {Wang}, \citenamefont {Nair}, \citenamefont {Kaiser}, \citenamefont {Wu}, \citenamefont {Geim},\ and\ \citenamefont {Grigorieva}}]{Algara_Siller_2015}%
  \BibitemOpen
  \bibfield  {author} {\bibinfo {author} {\bibfnamefont {G.}~\bibnamefont {Algara-Siller}}, \bibinfo {author} {\bibfnamefont {O.}~\bibnamefont {Lehtinen}}, \bibinfo {author} {\bibfnamefont {F.~C.}\ \bibnamefont {Wang}}, \bibinfo {author} {\bibfnamefont {R.~R.}\ \bibnamefont {Nair}}, \bibinfo {author} {\bibfnamefont {U.}~\bibnamefont {Kaiser}}, \bibinfo {author} {\bibfnamefont {H.~A.}\ \bibnamefont {Wu}}, \bibinfo {author} {\bibfnamefont {A.~K.}\ \bibnamefont {Geim}},\ and\ \bibinfo {author} {\bibfnamefont {I.~V.}\ \bibnamefont {Grigorieva}},\ }\href {https://doi.org/10.1038/nature14295} {\bibfield  {journal} {\bibinfo  {journal} {Nature}\ }\textbf {\bibinfo {volume} {519}},\ \bibinfo {pages} {443–445} (\bibinfo {year} {2015})}\BibitemShut {NoStop}%
\bibitem [{\citenamefont {Wang}\ \emph {et~al.}(2025)\citenamefont {Wang}, \citenamefont {Souilamas}, \citenamefont {Esfandiar}, \citenamefont {Fabregas}, \citenamefont {Benaglia}, \citenamefont {Nevison-Andrews}, \citenamefont {Yang}, \citenamefont {Normansell}, \citenamefont {Ares}, \citenamefont {Ferrari}, \citenamefont {Principi}, \citenamefont {Geim},\ and\ \citenamefont {Fumagalli}}]{wang2025inplane}%
  \BibitemOpen
  \bibfield  {author} {\bibinfo {author} {\bibfnamefont {R.}~\bibnamefont {Wang}}, \bibinfo {author} {\bibfnamefont {M.}~\bibnamefont {Souilamas}}, \bibinfo {author} {\bibfnamefont {A.}~\bibnamefont {Esfandiar}}, \bibinfo {author} {\bibfnamefont {R.}~\bibnamefont {Fabregas}}, \bibinfo {author} {\bibfnamefont {S.}~\bibnamefont {Benaglia}}, \bibinfo {author} {\bibfnamefont {H.}~\bibnamefont {Nevison-Andrews}}, \bibinfo {author} {\bibfnamefont {Q.}~\bibnamefont {Yang}}, \bibinfo {author} {\bibfnamefont {J.}~\bibnamefont {Normansell}}, \bibinfo {author} {\bibfnamefont {P.}~\bibnamefont {Ares}}, \bibinfo {author} {\bibfnamefont {G.}~\bibnamefont {Ferrari}}, \bibinfo {author} {\bibfnamefont {A.}~\bibnamefont {Principi}}, \bibinfo {author} {\bibfnamefont {A.~K.}\ \bibnamefont {Geim}},\ and\ \bibinfo {author} {\bibfnamefont {L.}~\bibnamefont {Fumagalli}},\ }\Eprint {https://arxiv.org/abs/2407.21538} {arXiv:2407.21538}  (\bibinfo {year} {2025}),\ \bibinfo {note} {arXiv preprint arXiv:2407.21538}\BibitemShut {NoStop}%
\bibitem [{\citenamefont {Jiang}\ \emph {et~al.}(2024)\citenamefont {Jiang}, \citenamefont {Gao}, \citenamefont {Li}, \citenamefont {Liu}, \citenamefont {Zhu}, \citenamefont {Zhu}, \citenamefont {Francisco},\ and\ \citenamefont {Zeng}}]{Jiang_2024}%
  \BibitemOpen
  \bibfield  {author} {\bibinfo {author} {\bibfnamefont {J.}~\bibnamefont {Jiang}}, \bibinfo {author} {\bibfnamefont {Y.}~\bibnamefont {Gao}}, \bibinfo {author} {\bibfnamefont {L.}~\bibnamefont {Li}}, \bibinfo {author} {\bibfnamefont {Y.}~\bibnamefont {Liu}}, \bibinfo {author} {\bibfnamefont {W.}~\bibnamefont {Zhu}}, \bibinfo {author} {\bibfnamefont {C.}~\bibnamefont {Zhu}}, \bibinfo {author} {\bibfnamefont {J.~S.}\ \bibnamefont {Francisco}},\ and\ \bibinfo {author} {\bibfnamefont {X.~C.}\ \bibnamefont {Zeng}},\ }\href {https://doi.org/10.1038/s41567-023-02341-8} {\bibfield  {journal} {\bibinfo  {journal} {Nature Physics}\ }\textbf {\bibinfo {volume} {20}},\ \bibinfo {pages} {456–464} (\bibinfo {year} {2024})}\BibitemShut {NoStop}%
\bibitem [{\citenamefont {Ravindra}\ \emph {et~al.}(2024{\natexlab{a}})\citenamefont {Ravindra}, \citenamefont {Advincula}, \citenamefont {Shi}, \citenamefont {Coles}, \citenamefont {Michaelides},\ and\ \citenamefont {Kapil}}]{ravindra2024nuclearquantumeffectsinduce}%
  \BibitemOpen
  \bibfield  {author} {\bibinfo {author} {\bibfnamefont {P.}~\bibnamefont {Ravindra}}, \bibinfo {author} {\bibfnamefont {X.~R.}\ \bibnamefont {Advincula}}, \bibinfo {author} {\bibfnamefont {B.~X.}\ \bibnamefont {Shi}}, \bibinfo {author} {\bibfnamefont {S.~W.}\ \bibnamefont {Coles}}, \bibinfo {author} {\bibfnamefont {A.}~\bibnamefont {Michaelides}},\ and\ \bibinfo {author} {\bibfnamefont {V.}~\bibnamefont {Kapil}},\ }\href {https://arxiv.org/abs/2410.03272} {\bibfield  {journal} {\bibinfo  {journal} {arXiv preprint arXiv:2410.03272}\ } (\bibinfo {year} {2024}{\natexlab{a}})}\BibitemShut {NoStop}%
\bibitem [{\citenamefont {Laage}\ and\ \citenamefont {Hynes}(2006)}]{Laage_2006}%
  \BibitemOpen
  \bibfield  {author} {\bibinfo {author} {\bibfnamefont {D.}~\bibnamefont {Laage}}\ and\ \bibinfo {author} {\bibfnamefont {J.~T.}\ \bibnamefont {Hynes}},\ }\href {https://doi.org/10.1126/science.1122154} {\bibfield  {journal} {\bibinfo  {journal} {Science}\ }\textbf {\bibinfo {volume} {311}},\ \bibinfo {pages} {832–835} (\bibinfo {year} {2006})}\BibitemShut {NoStop}%
\bibitem [{\citenamefont {Hassanali}\ \emph {et~al.}(2013)\citenamefont {Hassanali}, \citenamefont {Giberti}, \citenamefont {Cuny}, \citenamefont {Kühne},\ and\ \citenamefont {Parrinello}}]{Hassanali_2013}%
  \BibitemOpen
  \bibfield  {author} {\bibinfo {author} {\bibfnamefont {A.}~\bibnamefont {Hassanali}}, \bibinfo {author} {\bibfnamefont {F.}~\bibnamefont {Giberti}}, \bibinfo {author} {\bibfnamefont {J.}~\bibnamefont {Cuny}}, \bibinfo {author} {\bibfnamefont {T.~D.}\ \bibnamefont {Kühne}},\ and\ \bibinfo {author} {\bibfnamefont {M.}~\bibnamefont {Parrinello}},\ }\href {https://doi.org/10.1073/pnas.1306642110} {\bibfield  {journal} {\bibinfo  {journal} {Proceedings of the National Academy of Sciences}\ }\textbf {\bibinfo {volume} {110}},\ \bibinfo {pages} {13723–13728} (\bibinfo {year} {2013})}\BibitemShut {NoStop}%
\bibitem [{\citenamefont {Bernal}\ and\ \citenamefont {Fowler}(1933)}]{Bernal_1933}%
  \BibitemOpen
  \bibfield  {author} {\bibinfo {author} {\bibfnamefont {J.~D.}\ \bibnamefont {Bernal}}\ and\ \bibinfo {author} {\bibfnamefont {R.~H.}\ \bibnamefont {Fowler}},\ }\href {https://doi.org/10.1063/1.1749327} {\bibfield  {journal} {\bibinfo  {journal} {J. Chem. Phys.}\ }\textbf {\bibinfo {volume} {1}},\ \bibinfo {pages} {515} (\bibinfo {year} {1933})}\BibitemShut {NoStop}%
\bibitem [{\citenamefont {Das}\ \emph {et~al.}(2024)\citenamefont {Das}, \citenamefont {Ruiz-Barragan}, \citenamefont {Bagchi},\ and\ \citenamefont {Marx}}]{Das_2024}%
  \BibitemOpen
  \bibfield  {author} {\bibinfo {author} {\bibfnamefont {B.}~\bibnamefont {Das}}, \bibinfo {author} {\bibfnamefont {S.}~\bibnamefont {Ruiz-Barragan}}, \bibinfo {author} {\bibfnamefont {B.}~\bibnamefont {Bagchi}},\ and\ \bibinfo {author} {\bibfnamefont {D.}~\bibnamefont {Marx}},\ }\href {https://doi.org/10.1021/acs.nanolett.4c04077} {\bibfield  {journal} {\bibinfo  {journal} {Nano Letters}\ }\textbf {\bibinfo {volume} {24}},\ \bibinfo {pages} {15623–15628} (\bibinfo {year} {2024})}\BibitemShut {NoStop}%
\bibitem [{\citenamefont {Sun}\ \emph {et~al.}(2015)\citenamefont {Sun}, \citenamefont {Clark}, \citenamefont {Torquato},\ and\ \citenamefont {Car}}]{Sun_2015}%
  \BibitemOpen
  \bibfield  {author} {\bibinfo {author} {\bibfnamefont {J.}~\bibnamefont {Sun}}, \bibinfo {author} {\bibfnamefont {B.~K.}\ \bibnamefont {Clark}}, \bibinfo {author} {\bibfnamefont {S.}~\bibnamefont {Torquato}},\ and\ \bibinfo {author} {\bibfnamefont {R.}~\bibnamefont {Car}},\ }\href {https://doi.org/10.1038/ncomms9156} {\bibfield  {journal} {\bibinfo  {journal} {Nat. Commun.}\ }\textbf {\bibinfo {volume} {6}},\ \bibinfo {pages} {8156} (\bibinfo {year} {2015})}\BibitemShut {NoStop}%
\bibitem [{\citenamefont {Morgan}\ and\ \citenamefont {Madden}(2004)}]{Morgan_2004}%
  \BibitemOpen
  \bibfield  {author} {\bibinfo {author} {\bibfnamefont {B.}~\bibnamefont {Morgan}}\ and\ \bibinfo {author} {\bibfnamefont {P.~A.}\ \bibnamefont {Madden}},\ }\href {https://doi.org/10.1063/1.1629076} {\bibfield  {journal} {\bibinfo  {journal} {J. Chem. Phys.}\ }\textbf {\bibinfo {volume} {120}},\ \bibinfo {pages} {1402} (\bibinfo {year} {2004})}\BibitemShut {NoStop}%
\bibitem [{\citenamefont {Simoes~Santos}\ \emph {et~al.}(2024)\citenamefont {Simoes~Santos}, \citenamefont {Salanne}, \citenamefont {Kooyman},\ and\ \citenamefont {Lambertin}}]{Simoes_Santos_2024}%
  \BibitemOpen
  \bibfield  {author} {\bibinfo {author} {\bibfnamefont {M.}~\bibnamefont {Simoes~Santos}}, \bibinfo {author} {\bibfnamefont {M.}~\bibnamefont {Salanne}}, \bibinfo {author} {\bibfnamefont {T.}~\bibnamefont {Kooyman}},\ and\ \bibinfo {author} {\bibfnamefont {D.}~\bibnamefont {Lambertin}},\ }\href {https://doi.org/10.1016/j.jnucmat.2024.155125} {\bibfield  {journal} {\bibinfo  {journal} {Journal of Nuclear Materials}\ }\textbf {\bibinfo {volume} {597}},\ \bibinfo {pages} {155125} (\bibinfo {year} {2024})}\BibitemShut {NoStop}%
\bibitem [{\citenamefont {Hull}(2004)}]{Hull_2004}%
  \BibitemOpen
  \bibfield  {author} {\bibinfo {author} {\bibfnamefont {S.}~\bibnamefont {Hull}},\ }\href {https://doi.org/10.1088/0034-4885/67/7/r05} {\bibfield  {journal} {\bibinfo  {journal} {Reports on Progress in Physics}\ }\textbf {\bibinfo {volume} {67}},\ \bibinfo {pages} {1233–1314} (\bibinfo {year} {2004})}\BibitemShut {NoStop}%
\bibitem [{\citenamefont {Müller}\ \emph {et~al.}(2021)\citenamefont {Müller}, \citenamefont {Ertural}, \citenamefont {Hempelmann},\ and\ \citenamefont {Dronskowski}}]{M_ller_2021}%
  \BibitemOpen
  \bibfield  {author} {\bibinfo {author} {\bibfnamefont {P.~C.}\ \bibnamefont {Müller}}, \bibinfo {author} {\bibfnamefont {C.}~\bibnamefont {Ertural}}, \bibinfo {author} {\bibfnamefont {J.}~\bibnamefont {Hempelmann}},\ and\ \bibinfo {author} {\bibfnamefont {R.}~\bibnamefont {Dronskowski}},\ }\href {https://doi.org/10.1021/acs.jpcc.1c00718} {\bibfield  {journal} {\bibinfo  {journal} {J. Phys. Chem. C}\ }\textbf {\bibinfo {volume} {125}},\ \bibinfo {pages} {7959} (\bibinfo {year} {2021})}\BibitemShut {NoStop}%
\bibitem [{\citenamefont {Ninet}\ \emph {et~al.}(2012)\citenamefont {Ninet}, \citenamefont {Datchi},\ and\ \citenamefont {Saitta}}]{Ninet_2012}%
  \BibitemOpen
  \bibfield  {author} {\bibinfo {author} {\bibfnamefont {S.}~\bibnamefont {Ninet}}, \bibinfo {author} {\bibfnamefont {F.}~\bibnamefont {Datchi}},\ and\ \bibinfo {author} {\bibfnamefont {A.~M.}\ \bibnamefont {Saitta}},\ }\href {https://doi.org/10.1103/PhysRevLett.108.165702} {\bibfield  {journal} {\bibinfo  {journal} {Phys. Rev. Lett.}\ }\textbf {\bibinfo {volume} {108}},\ \bibinfo {pages} {165702} (\bibinfo {year} {2012})}\BibitemShut {NoStop}%
\bibitem [{\citenamefont {Pickard}\ and\ \citenamefont {Needs}(2008)}]{Pickard_2008}%
  \BibitemOpen
  \bibfield  {author} {\bibinfo {author} {\bibfnamefont {C.~J.}\ \bibnamefont {Pickard}}\ and\ \bibinfo {author} {\bibfnamefont {R.~J.}\ \bibnamefont {Needs}},\ }\href {https://doi.org/10.1038/nmat2261} {\bibfield  {journal} {\bibinfo  {journal} {Nature Materials}\ }\textbf {\bibinfo {volume} {7}},\ \bibinfo {pages} {775} (\bibinfo {year} {2008})}\BibitemShut {NoStop}%
\bibitem [{\citenamefont {Kreuer}(1996)}]{Kreuer_1996}%
  \BibitemOpen
  \bibfield  {author} {\bibinfo {author} {\bibfnamefont {K.-D.}\ \bibnamefont {Kreuer}},\ }\href {https://doi.org/10.1021/cm950192a} {\bibfield  {journal} {\bibinfo  {journal} {Chemistry of Materials}\ }\textbf {\bibinfo {volume} {8}},\ \bibinfo {pages} {610–641} (\bibinfo {year} {1996})}\BibitemShut {NoStop}%
\bibitem [{\citenamefont {Wood}\ \emph {et~al.}(2021)\citenamefont {Wood}, \citenamefont {Varley}, \citenamefont {Kweon}, \citenamefont {Shea}, \citenamefont {Hall}, \citenamefont {Grieder}, \citenamefont {Ward}, \citenamefont {Aguirre}, \citenamefont {Rigling}, \citenamefont {Lopez~Ventura}, \citenamefont {Stancill},\ and\ \citenamefont {Adelstein}}]{Wood_2021}%
  \BibitemOpen
  \bibfield  {author} {\bibinfo {author} {\bibfnamefont {B.~C.}\ \bibnamefont {Wood}}, \bibinfo {author} {\bibfnamefont {J.~B.}\ \bibnamefont {Varley}}, \bibinfo {author} {\bibfnamefont {K.~E.}\ \bibnamefont {Kweon}}, \bibinfo {author} {\bibfnamefont {P.}~\bibnamefont {Shea}}, \bibinfo {author} {\bibfnamefont {A.~T.}\ \bibnamefont {Hall}}, \bibinfo {author} {\bibfnamefont {A.}~\bibnamefont {Grieder}}, \bibinfo {author} {\bibfnamefont {M.}~\bibnamefont {Ward}}, \bibinfo {author} {\bibfnamefont {V.~P.}\ \bibnamefont {Aguirre}}, \bibinfo {author} {\bibfnamefont {D.}~\bibnamefont {Rigling}}, \bibinfo {author} {\bibfnamefont {E.}~\bibnamefont {Lopez~Ventura}}, \bibinfo {author} {\bibfnamefont {C.}~\bibnamefont {Stancill}},\ and\ \bibinfo {author} {\bibfnamefont {N.}~\bibnamefont {Adelstein}},\ }\bibfield  {journal} {\bibinfo  {journal} {Philosophical Transactions of the Royal Society A: Mathematical, Physical and Engineering Sciences}\ }\textbf {\bibinfo {volume} {379}},\ \href
  {https://doi.org/10.1098/rsta.2019.0467} {10.1098/rsta.2019.0467} (\bibinfo {year} {2021})\BibitemShut {NoStop}%
\bibitem [{\citenamefont {Catlow}(1990)}]{Catlow_1990}%
  \BibitemOpen
  \bibfield  {author} {\bibinfo {author} {\bibfnamefont {C.~R.~A.}\ \bibnamefont {Catlow}},\ }\href {https://doi.org/10.1039/ft9908601167} {\bibfield  {journal} {\bibinfo  {journal} {Journal of the Chemical Society, Faraday Transactions}\ }\textbf {\bibinfo {volume} {86}},\ \bibinfo {pages} {1167} (\bibinfo {year} {1990})}\BibitemShut {NoStop}%
\bibitem [{\citenamefont {Morgan}\ and\ \citenamefont {Madden}(2014)}]{Morgan_2014}%
  \BibitemOpen
  \bibfield  {author} {\bibinfo {author} {\bibfnamefont {B.~J.}\ \bibnamefont {Morgan}}\ and\ \bibinfo {author} {\bibfnamefont {P.~A.}\ \bibnamefont {Madden}},\ }\href {https://doi.org/10.1103/PhysRevLett.112.145901} {\bibfield  {journal} {\bibinfo  {journal} {Phys. Rev. Lett.}\ }\textbf {\bibinfo {volume} {112}},\ \bibinfo {pages} {145901} (\bibinfo {year} {2014})}\BibitemShut {NoStop}%
\bibitem [{\citenamefont {Joos}\ \emph {et~al.}(2025)\citenamefont {Joos}, \citenamefont {Kang}, \citenamefont {Merkle},\ and\ \citenamefont {Maier}}]{Joos_2025}%
  \BibitemOpen
  \bibfield  {author} {\bibinfo {author} {\bibfnamefont {M.}~\bibnamefont {Joos}}, \bibinfo {author} {\bibfnamefont {X.}~\bibnamefont {Kang}}, \bibinfo {author} {\bibfnamefont {R.}~\bibnamefont {Merkle}},\ and\ \bibinfo {author} {\bibfnamefont {J.}~\bibnamefont {Maier}},\ }\href {https://doi.org/10.1038/s41563-025-02143-8} {\bibfield  {journal} {\bibinfo  {journal} {Nat. Mater.}\ }\textbf {\bibinfo {volume} {24}},\ \bibinfo {pages} {397} (\bibinfo {year} {2025})}\BibitemShut {NoStop}%
\bibitem [{\citenamefont {de~Grotthuss}(1805)}]{de1805moire}%
  \BibitemOpen
  \bibfield  {author} {\bibinfo {author} {\bibfnamefont {C.}~\bibnamefont {de~Grotthuss}},\ }\href {https://books.google.co.uk/books?id=ORxIjwEACAAJ} {\emph {\bibinfo {title} {M{\'e}moire sur la d{\'e}composition de l'eau: et des corps qu' elle tient en dissolution {\`a} l'aide de l'{\'e}lectricit{\'e} galvanique}}}\ (\bibinfo {year} {1805})\BibitemShut {NoStop}%
\bibitem [{\citenamefont {Marx}(2006)}]{Marx_2006}%
  \BibitemOpen
  \bibfield  {author} {\bibinfo {author} {\bibfnamefont {D.}~\bibnamefont {Marx}},\ }\href {https://doi.org/10.1002/cphc.200600128} {\bibfield  {journal} {\bibinfo  {journal} {ChemPhysChem}\ }\textbf {\bibinfo {volume} {7}},\ \bibinfo {pages} {1848–1870} (\bibinfo {year} {2006})}\BibitemShut {NoStop}%
\bibitem [{\citenamefont {Chen}\ \emph {et~al.}(2018)\citenamefont {Chen}, \citenamefont {Zheng}, \citenamefont {Santra}, \citenamefont {Ko}, \citenamefont {DiStasio~Jr}, \citenamefont {Klein}, \citenamefont {Car},\ and\ \citenamefont {Wu}}]{Chen_2018}%
  \BibitemOpen
  \bibfield  {author} {\bibinfo {author} {\bibfnamefont {M.}~\bibnamefont {Chen}}, \bibinfo {author} {\bibfnamefont {L.}~\bibnamefont {Zheng}}, \bibinfo {author} {\bibfnamefont {B.}~\bibnamefont {Santra}}, \bibinfo {author} {\bibfnamefont {H.-Y.}\ \bibnamefont {Ko}}, \bibinfo {author} {\bibfnamefont {R.~A.}\ \bibnamefont {DiStasio~Jr}}, \bibinfo {author} {\bibfnamefont {M.~L.}\ \bibnamefont {Klein}}, \bibinfo {author} {\bibfnamefont {R.}~\bibnamefont {Car}},\ and\ \bibinfo {author} {\bibfnamefont {X.}~\bibnamefont {Wu}},\ }\href {https://doi.org/10.1038/s41557-018-0010-2} {\bibfield  {journal} {\bibinfo  {journal} {Nature Chemistry}\ }\textbf {\bibinfo {volume} {10}},\ \bibinfo {pages} {413–419} (\bibinfo {year} {2018})}\BibitemShut {NoStop}%
\bibitem [{\citenamefont {Tuckerman}\ \emph {et~al.}(2006)\citenamefont {Tuckerman}, \citenamefont {Chandra},\ and\ \citenamefont {Marx}}]{Tuckerman_2006}%
  \BibitemOpen
  \bibfield  {author} {\bibinfo {author} {\bibfnamefont {M.~E.}\ \bibnamefont {Tuckerman}}, \bibinfo {author} {\bibfnamefont {A.}~\bibnamefont {Chandra}},\ and\ \bibinfo {author} {\bibfnamefont {D.}~\bibnamefont {Marx}},\ }\href {https://doi.org/10.1021/ar040207n} {\bibfield  {journal} {\bibinfo  {journal} {Accounts of Chemical Research}\ }\textbf {\bibinfo {volume} {39}},\ \bibinfo {pages} {151–158} (\bibinfo {year} {2006})}\BibitemShut {NoStop}%
\bibitem [{\citenamefont {Marx}\ \emph {et~al.}(1999)\citenamefont {Marx}, \citenamefont {Tuckerman}, \citenamefont {Hutter},\ and\ \citenamefont {Parrinello}}]{Marx_1999}%
  \BibitemOpen
  \bibfield  {author} {\bibinfo {author} {\bibfnamefont {D.}~\bibnamefont {Marx}}, \bibinfo {author} {\bibfnamefont {M.~E.}\ \bibnamefont {Tuckerman}}, \bibinfo {author} {\bibfnamefont {J.}~\bibnamefont {Hutter}},\ and\ \bibinfo {author} {\bibfnamefont {M.}~\bibnamefont {Parrinello}},\ }\href {https://doi.org/10.1038/17579} {\bibfield  {journal} {\bibinfo  {journal} {Nature}\ }\textbf {\bibinfo {volume} {397}},\ \bibinfo {pages} {601–604} (\bibinfo {year} {1999})}\BibitemShut {NoStop}%
\bibitem [{\citenamefont {Burbano}\ \emph {et~al.}(2016)\citenamefont {Burbano}, \citenamefont {Carlier}, \citenamefont {Boucher}, \citenamefont {Morgan},\ and\ \citenamefont {Salanne}}]{Burbano_2016}%
  \BibitemOpen
  \bibfield  {author} {\bibinfo {author} {\bibfnamefont {M.}~\bibnamefont {Burbano}}, \bibinfo {author} {\bibfnamefont {D.}~\bibnamefont {Carlier}}, \bibinfo {author} {\bibfnamefont {F.}~\bibnamefont {Boucher}}, \bibinfo {author} {\bibfnamefont {B.~J.}\ \bibnamefont {Morgan}},\ and\ \bibinfo {author} {\bibfnamefont {M.}~\bibnamefont {Salanne}},\ }\href {https://doi.org/10.1103/PhysRevLett.116.135901} {\bibfield  {journal} {\bibinfo  {journal} {Phys. Rev. Lett.}\ }\textbf {\bibinfo {volume} {116}},\ \bibinfo {pages} {135901} (\bibinfo {year} {2016})}\BibitemShut {NoStop}%
\bibitem [{\citenamefont {Morgan}(2021{\natexlab{a}})}]{Morgan_2021}%
  \BibitemOpen
  \bibfield  {author} {\bibinfo {author} {\bibfnamefont {B.~J.}\ \bibnamefont {Morgan}},\ }\href {https://doi.org/10.1021/acs.chemmater.0c03738} {\bibfield  {journal} {\bibinfo  {journal} {Chemistry of Materials}\ }\textbf {\bibinfo {volume} {33}},\ \bibinfo {pages} {2004–2018} (\bibinfo {year} {2021}{\natexlab{a}})}\BibitemShut {NoStop}%
\bibitem [{\citenamefont {Annamareddy}\ and\ \citenamefont {Eapen}(2017)}]{Annamareddy_2017}%
  \BibitemOpen
  \bibfield  {author} {\bibinfo {author} {\bibfnamefont {A.}~\bibnamefont {Annamareddy}}\ and\ \bibinfo {author} {\bibfnamefont {J.}~\bibnamefont {Eapen}},\ }\href {https://doi.org/10.1038/srep44149} {\bibfield  {journal} {\bibinfo  {journal} {Sci. Rep.}\ }\textbf {\bibinfo {volume} {7}},\ \bibinfo {pages} {44149} (\bibinfo {year} {2017})}\BibitemShut {NoStop}%
\bibitem [{\citenamefont {Futera}\ \emph {et~al.}(2020)\citenamefont {Futera}, \citenamefont {Tse},\ and\ \citenamefont {English}}]{Futera_2020}%
  \BibitemOpen
  \bibfield  {author} {\bibinfo {author} {\bibfnamefont {Z.}~\bibnamefont {Futera}}, \bibinfo {author} {\bibfnamefont {J.~S.}\ \bibnamefont {Tse}},\ and\ \bibinfo {author} {\bibfnamefont {N.~J.}\ \bibnamefont {English}},\ }\bibfield  {journal} {\bibinfo  {journal} {Science Advances}\ }\textbf {\bibinfo {volume} {6}},\ \href {https://doi.org/10.1126/sciadv.aaz2915} {10.1126/sciadv.aaz2915} (\bibinfo {year} {2020})\BibitemShut {NoStop}%
\bibitem [{\citenamefont {Noguchi}\ and\ \citenamefont {Okuchi}(2016)}]{Noguchi_2016}%
  \BibitemOpen
  \bibfield  {author} {\bibinfo {author} {\bibfnamefont {N.}~\bibnamefont {Noguchi}}\ and\ \bibinfo {author} {\bibfnamefont {T.}~\bibnamefont {Okuchi}},\ }\href {https://doi.org/10.1063/1.4953688} {\bibfield  {journal} {\bibinfo  {journal} {J. Chem. Phys.}\ }\textbf {\bibinfo {volume} {144}},\ \bibinfo {pages} {234507} (\bibinfo {year} {2016})}\BibitemShut {NoStop}%
\bibitem [{\citenamefont {Li}\ \emph {et~al.}(2010)\citenamefont {Li}, \citenamefont {Probert}, \citenamefont {Alavi},\ and\ \citenamefont {Michaelides}}]{Li_2010}%
  \BibitemOpen
  \bibfield  {author} {\bibinfo {author} {\bibfnamefont {X.-Z.}\ \bibnamefont {Li}}, \bibinfo {author} {\bibfnamefont {M.~I.~J.}\ \bibnamefont {Probert}}, \bibinfo {author} {\bibfnamefont {A.}~\bibnamefont {Alavi}},\ and\ \bibinfo {author} {\bibfnamefont {A.}~\bibnamefont {Michaelides}},\ }\href {https://doi.org/10.1103/PhysRevLett.104.066102} {\bibfield  {journal} {\bibinfo  {journal} {Phys. Rev. Lett.}\ }\textbf {\bibinfo {volume} {104}},\ \bibinfo {pages} {066102} (\bibinfo {year} {2010})}\BibitemShut {NoStop}%
\bibitem [{\citenamefont {Gomez}\ \emph {et~al.}(2024)\citenamefont {Gomez}, \citenamefont {Thompson},\ and\ \citenamefont {Laage}}]{Gomez_2024}%
  \BibitemOpen
  \bibfield  {author} {\bibinfo {author} {\bibfnamefont {A.}~\bibnamefont {Gomez}}, \bibinfo {author} {\bibfnamefont {W.~H.}\ \bibnamefont {Thompson}},\ and\ \bibinfo {author} {\bibfnamefont {D.}~\bibnamefont {Laage}},\ }\href {https://doi.org/10.1038/s41557-024-01593-y} {\bibfield  {journal} {\bibinfo  {journal} {Nature Chemistry}\ }\textbf {\bibinfo {volume} {16}},\ \bibinfo {pages} {1838–1844} (\bibinfo {year} {2024})}\BibitemShut {NoStop}%
\bibitem [{\citenamefont {Chandra}\ \emph {et~al.}(2007)\citenamefont {Chandra}, \citenamefont {Tuckerman},\ and\ \citenamefont {Marx}}]{Chandra_2007}%
  \BibitemOpen
  \bibfield  {author} {\bibinfo {author} {\bibfnamefont {A.}~\bibnamefont {Chandra}}, \bibinfo {author} {\bibfnamefont {M.~E.}\ \bibnamefont {Tuckerman}},\ and\ \bibinfo {author} {\bibfnamefont {D.}~\bibnamefont {Marx}},\ }\href {https://doi.org/10.1103/PhysRevLett.99.145901} {\bibfield  {journal} {\bibinfo  {journal} {Phys. Rev. Lett.}\ }\textbf {\bibinfo {volume} {99}},\ \bibinfo {pages} {145901} (\bibinfo {year} {2007})}\BibitemShut {NoStop}%
\bibitem [{\citenamefont {Ravindra}\ \emph {et~al.}(2024{\natexlab{b}})\citenamefont {Ravindra}, \citenamefont {Advincula}, \citenamefont {Schran}, \citenamefont {Michaelides},\ and\ \citenamefont {Kapil}}]{Ravindra_2024}%
  \BibitemOpen
  \bibfield  {author} {\bibinfo {author} {\bibfnamefont {P.}~\bibnamefont {Ravindra}}, \bibinfo {author} {\bibfnamefont {X.~R.}\ \bibnamefont {Advincula}}, \bibinfo {author} {\bibfnamefont {C.}~\bibnamefont {Schran}}, \bibinfo {author} {\bibfnamefont {A.}~\bibnamefont {Michaelides}},\ and\ \bibinfo {author} {\bibfnamefont {V.}~\bibnamefont {Kapil}},\ }\href {https://doi.org/10.1038/s41467-024-51124-z} {\bibfield  {journal} {\bibinfo  {journal} {Nat. Commun.}\ }\textbf {\bibinfo {volume} {15}},\ \bibinfo {pages} {7301} (\bibinfo {year} {2024}{\natexlab{b}})}\BibitemShut {NoStop}%
\bibitem [{\citenamefont {Tocci}\ and\ \citenamefont {Michaelides}(2014)}]{Tocci_2014}%
  \BibitemOpen
  \bibfield  {author} {\bibinfo {author} {\bibfnamefont {G.}~\bibnamefont {Tocci}}\ and\ \bibinfo {author} {\bibfnamefont {A.}~\bibnamefont {Michaelides}},\ }\href {https://doi.org/10.1021/jz402646c} {\bibfield  {journal} {\bibinfo  {journal} {The Journal of Physical Chemistry Letters}\ }\textbf {\bibinfo {volume} {5}},\ \bibinfo {pages} {474–480} (\bibinfo {year} {2014})}\BibitemShut {NoStop}%
\bibitem [{\citenamefont {Schran}\ \emph {et~al.}(2021)\citenamefont {Schran}, \citenamefont {Thiemann}, \citenamefont {Rowe}, \citenamefont {Müller}, \citenamefont {Marsalek},\ and\ \citenamefont {Michaelides}}]{Schran_2021}%
  \BibitemOpen
  \bibfield  {author} {\bibinfo {author} {\bibfnamefont {C.}~\bibnamefont {Schran}}, \bibinfo {author} {\bibfnamefont {F.~L.}\ \bibnamefont {Thiemann}}, \bibinfo {author} {\bibfnamefont {P.}~\bibnamefont {Rowe}}, \bibinfo {author} {\bibfnamefont {E.~A.}\ \bibnamefont {Müller}}, \bibinfo {author} {\bibfnamefont {O.}~\bibnamefont {Marsalek}},\ and\ \bibinfo {author} {\bibfnamefont {A.}~\bibnamefont {Michaelides}},\ }\href {https://doi.org/10.1073/pnas.2110077118} {\bibfield  {journal} {\bibinfo  {journal} {Proc. Natl. Acad. Sci. U.S.A.}\ }\textbf {\bibinfo {volume} {118}},\ \bibinfo {pages} {e2110077118} (\bibinfo {year} {2021})}\BibitemShut {NoStop}%
\bibitem [{\citenamefont {Kapil}\ \emph {et~al.}(2019)\citenamefont {Kapil}, \citenamefont {Rossi}, \citenamefont {Marsalek}, \citenamefont {Petraglia}, \citenamefont {Litman}, \citenamefont {Spura}, \citenamefont {Cheng}, \citenamefont {Cuzzocrea}, \citenamefont {Meißner}, \citenamefont {Wilkins}, \citenamefont {Helfrecht}, \citenamefont {Juda}, \citenamefont {Bienvenue}, \citenamefont {Fang}, \citenamefont {Kessler}, \citenamefont {Poltavsky}, \citenamefont {Vandenbrande}, \citenamefont {Wieme}, \citenamefont {Corminboeuf}, \citenamefont {Kühne}, \citenamefont {Manolopoulos}, \citenamefont {Markland}, \citenamefont {Richardson}, \citenamefont {Tkatchenko}, \citenamefont {Tribello}, \citenamefont {Van~Speybroeck},\ and\ \citenamefont {Ceriotti}}]{Kapil_2019}%
  \BibitemOpen
  \bibfield  {author} {\bibinfo {author} {\bibfnamefont {V.}~\bibnamefont {Kapil}}, \bibinfo {author} {\bibfnamefont {M.}~\bibnamefont {Rossi}}, \bibinfo {author} {\bibfnamefont {O.}~\bibnamefont {Marsalek}}, \bibinfo {author} {\bibfnamefont {R.}~\bibnamefont {Petraglia}}, \bibinfo {author} {\bibfnamefont {Y.}~\bibnamefont {Litman}}, \bibinfo {author} {\bibfnamefont {T.}~\bibnamefont {Spura}}, \bibinfo {author} {\bibfnamefont {B.}~\bibnamefont {Cheng}}, \bibinfo {author} {\bibfnamefont {A.}~\bibnamefont {Cuzzocrea}}, \bibinfo {author} {\bibfnamefont {R.~H.}\ \bibnamefont {Meißner}}, \bibinfo {author} {\bibfnamefont {D.~M.}\ \bibnamefont {Wilkins}}, \bibinfo {author} {\bibfnamefont {B.~A.}\ \bibnamefont {Helfrecht}}, \bibinfo {author} {\bibfnamefont {P.}~\bibnamefont {Juda}}, \bibinfo {author} {\bibfnamefont {S.~P.}\ \bibnamefont {Bienvenue}}, \bibinfo {author} {\bibfnamefont {W.}~\bibnamefont {Fang}}, \bibinfo {author} {\bibfnamefont {J.}~\bibnamefont {Kessler}}, \bibinfo {author} {\bibfnamefont
  {I.}~\bibnamefont {Poltavsky}}, \bibinfo {author} {\bibfnamefont {S.}~\bibnamefont {Vandenbrande}}, \bibinfo {author} {\bibfnamefont {J.}~\bibnamefont {Wieme}}, \bibinfo {author} {\bibfnamefont {C.}~\bibnamefont {Corminboeuf}}, \bibinfo {author} {\bibfnamefont {T.~D.}\ \bibnamefont {Kühne}}, \bibinfo {author} {\bibfnamefont {D.~E.}\ \bibnamefont {Manolopoulos}}, \bibinfo {author} {\bibfnamefont {T.~E.}\ \bibnamefont {Markland}}, \bibinfo {author} {\bibfnamefont {J.~O.}\ \bibnamefont {Richardson}}, \bibinfo {author} {\bibfnamefont {A.}~\bibnamefont {Tkatchenko}}, \bibinfo {author} {\bibfnamefont {G.~A.}\ \bibnamefont {Tribello}}, \bibinfo {author} {\bibfnamefont {V.}~\bibnamefont {Van~Speybroeck}},\ and\ \bibinfo {author} {\bibfnamefont {M.}~\bibnamefont {Ceriotti}},\ }\href {https://doi.org/10.1016/j.cpc.2018.09.020} {\bibfield  {journal} {\bibinfo  {journal} {Computer Physics Communications}\ }\textbf {\bibinfo {volume} {236}},\ \bibinfo {pages} {214–223} (\bibinfo {year} {2019})}\BibitemShut {NoStop}%
\bibitem [{\citenamefont {Singraber}\ \emph {et~al.}(2019)\citenamefont {Singraber}, \citenamefont {Behler},\ and\ \citenamefont {Dellago}}]{Singraber_2019}%
  \BibitemOpen
  \bibfield  {author} {\bibinfo {author} {\bibfnamefont {A.}~\bibnamefont {Singraber}}, \bibinfo {author} {\bibfnamefont {J.}~\bibnamefont {Behler}},\ and\ \bibinfo {author} {\bibfnamefont {C.}~\bibnamefont {Dellago}},\ }\href {https://doi.org/10.1021/acs.jctc.8b00770} {\bibfield  {journal} {\bibinfo  {journal} {Journal of Chemical Theory and Computation}\ }\textbf {\bibinfo {volume} {15}},\ \bibinfo {pages} {1827–1840} (\bibinfo {year} {2019})}\BibitemShut {NoStop}%
\bibitem [{\citenamefont {Hjorth~Larsen}\ \emph {et~al.}(2017)\citenamefont {Hjorth~Larsen}, \citenamefont {Jørgen~Mortensen}, \citenamefont {Blomqvist}, \citenamefont {Castelli}, \citenamefont {Christensen}, \citenamefont {Dułak}, \citenamefont {Friis}, \citenamefont {Groves}, \citenamefont {Hammer}, \citenamefont {Hargus}, \citenamefont {Hermes}, \citenamefont {Jennings}, \citenamefont {Bjerre~Jensen}, \citenamefont {Kermode}, \citenamefont {Kitchin}, \citenamefont {Leonhard~Kolsbjerg}, \citenamefont {Kubal}, \citenamefont {Kaasbjerg}, \citenamefont {Lysgaard}, \citenamefont {Bergmann~Maronsson}, \citenamefont {Maxson}, \citenamefont {Olsen}, \citenamefont {Pastewka}, \citenamefont {Peterson}, \citenamefont {Rostgaard}, \citenamefont {Schiøtz}, \citenamefont {Schütt}, \citenamefont {Strange}, \citenamefont {Thygesen}, \citenamefont {Vegge}, \citenamefont {Vilhelmsen}, \citenamefont {Walter}, \citenamefont {Zeng},\ and\ \citenamefont {Jacobsen}}]{Hjorth_Larsen_2017}%
  \BibitemOpen
  \bibfield  {author} {\bibinfo {author} {\bibfnamefont {A.}~\bibnamefont {Hjorth~Larsen}}, \bibinfo {author} {\bibfnamefont {J.}~\bibnamefont {Jørgen~Mortensen}}, \bibinfo {author} {\bibfnamefont {J.}~\bibnamefont {Blomqvist}}, \bibinfo {author} {\bibfnamefont {I.~E.}\ \bibnamefont {Castelli}}, \bibinfo {author} {\bibfnamefont {R.}~\bibnamefont {Christensen}}, \bibinfo {author} {\bibfnamefont {M.}~\bibnamefont {Dułak}}, \bibinfo {author} {\bibfnamefont {J.}~\bibnamefont {Friis}}, \bibinfo {author} {\bibfnamefont {M.~N.}\ \bibnamefont {Groves}}, \bibinfo {author} {\bibfnamefont {B.}~\bibnamefont {Hammer}}, \bibinfo {author} {\bibfnamefont {C.}~\bibnamefont {Hargus}}, \bibinfo {author} {\bibfnamefont {E.~D.}\ \bibnamefont {Hermes}}, \bibinfo {author} {\bibfnamefont {P.~C.}\ \bibnamefont {Jennings}}, \bibinfo {author} {\bibfnamefont {P.}~\bibnamefont {Bjerre~Jensen}}, \bibinfo {author} {\bibfnamefont {J.}~\bibnamefont {Kermode}}, \bibinfo {author} {\bibfnamefont {J.~R.}\ \bibnamefont {Kitchin}}, \bibinfo
  {author} {\bibfnamefont {E.}~\bibnamefont {Leonhard~Kolsbjerg}}, \bibinfo {author} {\bibfnamefont {J.}~\bibnamefont {Kubal}}, \bibinfo {author} {\bibfnamefont {K.}~\bibnamefont {Kaasbjerg}}, \bibinfo {author} {\bibfnamefont {S.}~\bibnamefont {Lysgaard}}, \bibinfo {author} {\bibfnamefont {J.}~\bibnamefont {Bergmann~Maronsson}}, \bibinfo {author} {\bibfnamefont {T.}~\bibnamefont {Maxson}}, \bibinfo {author} {\bibfnamefont {T.}~\bibnamefont {Olsen}}, \bibinfo {author} {\bibfnamefont {L.}~\bibnamefont {Pastewka}}, \bibinfo {author} {\bibfnamefont {A.}~\bibnamefont {Peterson}}, \bibinfo {author} {\bibfnamefont {C.}~\bibnamefont {Rostgaard}}, \bibinfo {author} {\bibfnamefont {J.}~\bibnamefont {Schiøtz}}, \bibinfo {author} {\bibfnamefont {O.}~\bibnamefont {Schütt}}, \bibinfo {author} {\bibfnamefont {M.}~\bibnamefont {Strange}}, \bibinfo {author} {\bibfnamefont {K.~S.}\ \bibnamefont {Thygesen}}, \bibinfo {author} {\bibfnamefont {T.}~\bibnamefont {Vegge}}, \bibinfo {author} {\bibfnamefont {L.}~\bibnamefont
  {Vilhelmsen}}, \bibinfo {author} {\bibfnamefont {M.}~\bibnamefont {Walter}}, \bibinfo {author} {\bibfnamefont {Z.}~\bibnamefont {Zeng}},\ and\ \bibinfo {author} {\bibfnamefont {K.~W.}\ \bibnamefont {Jacobsen}},\ }\href {https://doi.org/10.1088/1361-648x/aa680e} {\bibfield  {journal} {\bibinfo  {journal} {Journal of Physics: Condensed Matter}\ }\textbf {\bibinfo {volume} {29}},\ \bibinfo {pages} {273002} (\bibinfo {year} {2017})}\BibitemShut {NoStop}%
\bibitem [{\citenamefont {Behler}\ and\ \citenamefont {Parrinello}(2007)}]{Behler_2007}%
  \BibitemOpen
  \bibfield  {author} {\bibinfo {author} {\bibfnamefont {J.}~\bibnamefont {Behler}}\ and\ \bibinfo {author} {\bibfnamefont {M.}~\bibnamefont {Parrinello}},\ }\href {https://doi.org/10.1103/PhysRevLett.98.146401} {\bibfield  {journal} {\bibinfo  {journal} {Phys. Rev. Lett.}\ }\textbf {\bibinfo {volume} {98}},\ \bibinfo {pages} {146401} (\bibinfo {year} {2007})}\BibitemShut {NoStop}%
\bibitem [{\citenamefont {Reinhardt}\ \emph {et~al.}(2022)\citenamefont {Reinhardt}, \citenamefont {Bethkenhagen}, \citenamefont {Coppari}, \citenamefont {Millot}, \citenamefont {Hamel},\ and\ \citenamefont {Cheng}}]{Reinhardt_2022}%
  \BibitemOpen
  \bibfield  {author} {\bibinfo {author} {\bibfnamefont {A.}~\bibnamefont {Reinhardt}}, \bibinfo {author} {\bibfnamefont {M.}~\bibnamefont {Bethkenhagen}}, \bibinfo {author} {\bibfnamefont {F.}~\bibnamefont {Coppari}}, \bibinfo {author} {\bibfnamefont {M.}~\bibnamefont {Millot}}, \bibinfo {author} {\bibfnamefont {S.}~\bibnamefont {Hamel}},\ and\ \bibinfo {author} {\bibfnamefont {B.}~\bibnamefont {Cheng}},\ }\href {https://doi.org/10.1038/s41467-022-32374-1} {\bibfield  {journal} {\bibinfo  {journal} {Nat. Commun.}\ }\textbf {\bibinfo {volume} {13}},\ \bibinfo {pages} {4707} (\bibinfo {year} {2022})}\BibitemShut {NoStop}%
\bibitem [{\citenamefont {Thompson}\ \emph {et~al.}(2022)\citenamefont {Thompson}, \citenamefont {Aktulga}, \citenamefont {Berger}, \citenamefont {Bolintineanu}, \citenamefont {Brown}, \citenamefont {Crozier}, \citenamefont {in~’t Veld}, \citenamefont {Kohlmeyer}, \citenamefont {Moore}, \citenamefont {Nguyen}, \citenamefont {Shan}, \citenamefont {Stevens}, \citenamefont {Tranchida}, \citenamefont {Trott},\ and\ \citenamefont {Plimpton}}]{Thompson_2022}%
  \BibitemOpen
  \bibfield  {author} {\bibinfo {author} {\bibfnamefont {A.~P.}\ \bibnamefont {Thompson}}, \bibinfo {author} {\bibfnamefont {H.~M.}\ \bibnamefont {Aktulga}}, \bibinfo {author} {\bibfnamefont {R.}~\bibnamefont {Berger}}, \bibinfo {author} {\bibfnamefont {D.~S.}\ \bibnamefont {Bolintineanu}}, \bibinfo {author} {\bibfnamefont {W.~M.}\ \bibnamefont {Brown}}, \bibinfo {author} {\bibfnamefont {P.~S.}\ \bibnamefont {Crozier}}, \bibinfo {author} {\bibfnamefont {P.~J.}\ \bibnamefont {in~’t Veld}}, \bibinfo {author} {\bibfnamefont {A.}~\bibnamefont {Kohlmeyer}}, \bibinfo {author} {\bibfnamefont {S.~G.}\ \bibnamefont {Moore}}, \bibinfo {author} {\bibfnamefont {T.~D.}\ \bibnamefont {Nguyen}}, \bibinfo {author} {\bibfnamefont {R.}~\bibnamefont {Shan}}, \bibinfo {author} {\bibfnamefont {M.~J.}\ \bibnamefont {Stevens}}, \bibinfo {author} {\bibfnamefont {J.}~\bibnamefont {Tranchida}}, \bibinfo {author} {\bibfnamefont {C.}~\bibnamefont {Trott}},\ and\ \bibinfo {author} {\bibfnamefont {S.~J.}\ \bibnamefont {Plimpton}},\
  }\href {https://doi.org/10.1016/j.cpc.2021.108171} {\bibfield  {journal} {\bibinfo  {journal} {Computer Physics Communications}\ }\textbf {\bibinfo {volume} {271}},\ \bibinfo {pages} {108171} (\bibinfo {year} {2022})}\BibitemShut {NoStop}%
\bibitem [{\citenamefont {Kresse}\ and\ \citenamefont {Hafner}(1994)}]{Kresse_1994}%
  \BibitemOpen
  \bibfield  {author} {\bibinfo {author} {\bibfnamefont {G.}~\bibnamefont {Kresse}}\ and\ \bibinfo {author} {\bibfnamefont {J.}~\bibnamefont {Hafner}},\ }\href {https://doi.org/10.1103/physrevb.49.14251} {\bibfield  {journal} {\bibinfo  {journal} {Physical Review B}\ }\textbf {\bibinfo {volume} {49}},\ \bibinfo {pages} {14251–14269} (\bibinfo {year} {1994})}\BibitemShut {NoStop}%
\bibitem [{\citenamefont {Kresse}\ and\ \citenamefont {Furthmüller}(1996{\natexlab{a}})}]{Kresse_1996}%
  \BibitemOpen
  \bibfield  {author} {\bibinfo {author} {\bibfnamefont {G.}~\bibnamefont {Kresse}}\ and\ \bibinfo {author} {\bibfnamefont {J.}~\bibnamefont {Furthmüller}},\ }\href {https://doi.org/10.1103/physrevb.54.11169} {\bibfield  {journal} {\bibinfo  {journal} {Physical Review B}\ }\textbf {\bibinfo {volume} {54}},\ \bibinfo {pages} {11169–11186} (\bibinfo {year} {1996}{\natexlab{a}})}\BibitemShut {NoStop}%
\bibitem [{\citenamefont {Kresse}\ and\ \citenamefont {Furthmüller}(1996{\natexlab{b}})}]{KRESSE199615}%
  \BibitemOpen
  \bibfield  {author} {\bibinfo {author} {\bibfnamefont {G.}~\bibnamefont {Kresse}}\ and\ \bibinfo {author} {\bibfnamefont {J.}~\bibnamefont {Furthmüller}},\ }\href {https://doi.org/https://doi.org/10.1016/0927-0256(96)00008-0} {\bibfield  {journal} {\bibinfo  {journal} {Computational Materials Science}\ }\textbf {\bibinfo {volume} {6}},\ \bibinfo {pages} {15} (\bibinfo {year} {1996}{\natexlab{b}})}\BibitemShut {NoStop}%
\bibitem [{\citenamefont {Perdew}\ \emph {et~al.}(1996)\citenamefont {Perdew}, \citenamefont {Burke},\ and\ \citenamefont {Ernzerhof}}]{Perdew_1996}%
  \BibitemOpen
  \bibfield  {author} {\bibinfo {author} {\bibfnamefont {J.~P.}\ \bibnamefont {Perdew}}, \bibinfo {author} {\bibfnamefont {K.}~\bibnamefont {Burke}},\ and\ \bibinfo {author} {\bibfnamefont {M.}~\bibnamefont {Ernzerhof}},\ }\href {https://doi.org/10.1103/physrevlett.77.3865} {\bibfield  {journal} {\bibinfo  {journal} {Physical Review Letters}\ }\textbf {\bibinfo {volume} {77}},\ \bibinfo {pages} {3865–3868} (\bibinfo {year} {1996})}\BibitemShut {NoStop}%
\bibitem [{\citenamefont {Nelson}\ \emph {et~al.}(2020)\citenamefont {Nelson}, \citenamefont {Ertural}, \citenamefont {George}, \citenamefont {Deringer}, \citenamefont {Hautier},\ and\ \citenamefont {Dronskowski}}]{Nelson_2020}%
  \BibitemOpen
  \bibfield  {author} {\bibinfo {author} {\bibfnamefont {R.}~\bibnamefont {Nelson}}, \bibinfo {author} {\bibfnamefont {C.}~\bibnamefont {Ertural}}, \bibinfo {author} {\bibfnamefont {J.}~\bibnamefont {George}}, \bibinfo {author} {\bibfnamefont {V.~L.}\ \bibnamefont {Deringer}}, \bibinfo {author} {\bibfnamefont {G.}~\bibnamefont {Hautier}},\ and\ \bibinfo {author} {\bibfnamefont {R.}~\bibnamefont {Dronskowski}},\ }\href {https://doi.org/10.1002/jcc.26353} {\bibfield  {journal} {\bibinfo  {journal} {J. Comput. Chem.}\ }\textbf {\bibinfo {volume} {41}},\ \bibinfo {pages} {1931} (\bibinfo {year} {2020})}\BibitemShut {NoStop}%
\bibitem [{\citenamefont {Ong}\ \emph {et~al.}(2013)\citenamefont {Ong}, \citenamefont {Richards}, \citenamefont {Jain}, \citenamefont {Hautier}, \citenamefont {Kocher}, \citenamefont {Cholia}, \citenamefont {Gunter}, \citenamefont {Chevrier}, \citenamefont {Persson},\ and\ \citenamefont {Ceder}}]{Ong_2013}%
  \BibitemOpen
  \bibfield  {author} {\bibinfo {author} {\bibfnamefont {S.~P.}\ \bibnamefont {Ong}}, \bibinfo {author} {\bibfnamefont {W.~D.}\ \bibnamefont {Richards}}, \bibinfo {author} {\bibfnamefont {A.}~\bibnamefont {Jain}}, \bibinfo {author} {\bibfnamefont {G.}~\bibnamefont {Hautier}}, \bibinfo {author} {\bibfnamefont {M.}~\bibnamefont {Kocher}}, \bibinfo {author} {\bibfnamefont {S.}~\bibnamefont {Cholia}}, \bibinfo {author} {\bibfnamefont {D.}~\bibnamefont {Gunter}}, \bibinfo {author} {\bibfnamefont {V.~L.}\ \bibnamefont {Chevrier}}, \bibinfo {author} {\bibfnamefont {K.~A.}\ \bibnamefont {Persson}},\ and\ \bibinfo {author} {\bibfnamefont {G.}~\bibnamefont {Ceder}},\ }\href {https://doi.org/10.1016/j.commatsci.2012.10.028} {\bibfield  {journal} {\bibinfo  {journal} {Computational Materials Science}\ }\textbf {\bibinfo {volume} {68}},\ \bibinfo {pages} {314–319} (\bibinfo {year} {2013})}\BibitemShut {NoStop}%
\bibitem [{\citenamefont {Morgan}(2021{\natexlab{b}})}]{Morgan_Vasppy}%
  \BibitemOpen
  \bibfield  {author} {\bibinfo {author} {\bibfnamefont {B.}~\bibnamefont {Morgan}},\ }\href@noop {} {\bibinfo {title} {{vasppy} (version 0.7.1.0) [software]}},\ \bibinfo {howpublished} {\url{https://pypi.org/project/vasppy/0.7.1.0/}} (\bibinfo {year} {2021}{\natexlab{b}}),\ \bibinfo {note} {python package}\BibitemShut {NoStop}%
\bibitem [{\citenamefont {Harris}\ \emph {et~al.}(2020)\citenamefont {Harris}, \citenamefont {Millman}, \citenamefont {van~der Walt}, \citenamefont {Gommers}, \citenamefont {Virtanen}, \citenamefont {Cournapeau}, \citenamefont {Wieser}, \citenamefont {Taylor}, \citenamefont {Berg}, \citenamefont {Smith}, \citenamefont {Kern}, \citenamefont {Picus}, \citenamefont {Hoyer}, \citenamefont {van Kerkwijk}, \citenamefont {Brett}, \citenamefont {Haldane}, \citenamefont {del Río}, \citenamefont {Wiebe}, \citenamefont {Peterson}, \citenamefont {Gérard-Marchant}, \citenamefont {Sheppard}, \citenamefont {Reddy}, \citenamefont {Weckesser}, \citenamefont {Abbasi}, \citenamefont {Gohlke},\ and\ \citenamefont {Oliphant}}]{Harris_2020}%
  \BibitemOpen
  \bibfield  {author} {\bibinfo {author} {\bibfnamefont {C.~R.}\ \bibnamefont {Harris}}, \bibinfo {author} {\bibfnamefont {K.~J.}\ \bibnamefont {Millman}}, \bibinfo {author} {\bibfnamefont {S.~J.}\ \bibnamefont {van~der Walt}}, \bibinfo {author} {\bibfnamefont {R.}~\bibnamefont {Gommers}}, \bibinfo {author} {\bibfnamefont {P.}~\bibnamefont {Virtanen}}, \bibinfo {author} {\bibfnamefont {D.}~\bibnamefont {Cournapeau}}, \bibinfo {author} {\bibfnamefont {E.}~\bibnamefont {Wieser}}, \bibinfo {author} {\bibfnamefont {J.}~\bibnamefont {Taylor}}, \bibinfo {author} {\bibfnamefont {S.}~\bibnamefont {Berg}}, \bibinfo {author} {\bibfnamefont {N.~J.}\ \bibnamefont {Smith}}, \bibinfo {author} {\bibfnamefont {R.}~\bibnamefont {Kern}}, \bibinfo {author} {\bibfnamefont {M.}~\bibnamefont {Picus}}, \bibinfo {author} {\bibfnamefont {S.}~\bibnamefont {Hoyer}}, \bibinfo {author} {\bibfnamefont {M.~H.}\ \bibnamefont {van Kerkwijk}}, \bibinfo {author} {\bibfnamefont {M.}~\bibnamefont {Brett}}, \bibinfo {author} {\bibfnamefont
  {A.}~\bibnamefont {Haldane}}, \bibinfo {author} {\bibfnamefont {J.~F.}\ \bibnamefont {del Río}}, \bibinfo {author} {\bibfnamefont {M.}~\bibnamefont {Wiebe}}, \bibinfo {author} {\bibfnamefont {P.}~\bibnamefont {Peterson}}, \bibinfo {author} {\bibfnamefont {P.}~\bibnamefont {Gérard-Marchant}}, \bibinfo {author} {\bibfnamefont {K.}~\bibnamefont {Sheppard}}, \bibinfo {author} {\bibfnamefont {T.}~\bibnamefont {Reddy}}, \bibinfo {author} {\bibfnamefont {W.}~\bibnamefont {Weckesser}}, \bibinfo {author} {\bibfnamefont {H.}~\bibnamefont {Abbasi}}, \bibinfo {author} {\bibfnamefont {C.}~\bibnamefont {Gohlke}},\ and\ \bibinfo {author} {\bibfnamefont {T.~E.}\ \bibnamefont {Oliphant}},\ }\href {https://doi.org/10.1038/s41586-020-2649-2} {\bibfield  {journal} {\bibinfo  {journal} {Nature}\ }\textbf {\bibinfo {volume} {585}},\ \bibinfo {pages} {357–362} (\bibinfo {year} {2020})}\BibitemShut {NoStop}%
\bibitem [{\citenamefont {Fong}\ \emph {et~al.}(2024)\citenamefont {Fong}, \citenamefont {Sumić}, \citenamefont {O’Neill}, \citenamefont {Schran}, \citenamefont {Grey},\ and\ \citenamefont {Michaelides}}]{Fong_2024}%
  \BibitemOpen
  \bibfield  {author} {\bibinfo {author} {\bibfnamefont {K.~D.}\ \bibnamefont {Fong}}, \bibinfo {author} {\bibfnamefont {B.}~\bibnamefont {Sumić}}, \bibinfo {author} {\bibfnamefont {N.}~\bibnamefont {O’Neill}}, \bibinfo {author} {\bibfnamefont {C.}~\bibnamefont {Schran}}, \bibinfo {author} {\bibfnamefont {C.~P.}\ \bibnamefont {Grey}},\ and\ \bibinfo {author} {\bibfnamefont {A.}~\bibnamefont {Michaelides}},\ }\href {https://doi.org/10.1021/acs.nanolett.4c00890} {\bibfield  {journal} {\bibinfo  {journal} {Nano Letters}\ }\textbf {\bibinfo {volume} {24}},\ \bibinfo {pages} {5024} (\bibinfo {year} {2024})}\BibitemShut {NoStop}%
\bibitem [{\citenamefont {Donati}\ \emph {et~al.}(1998)\citenamefont {Donati}, \citenamefont {Douglas}, \citenamefont {Kob}, \citenamefont {Plimpton}, \citenamefont {Poole},\ and\ \citenamefont {Glotzer}}]{Donati_1998}%
  \BibitemOpen
  \bibfield  {author} {\bibinfo {author} {\bibfnamefont {C.}~\bibnamefont {Donati}}, \bibinfo {author} {\bibfnamefont {J.~F.}\ \bibnamefont {Douglas}}, \bibinfo {author} {\bibfnamefont {W.}~\bibnamefont {Kob}}, \bibinfo {author} {\bibfnamefont {S.~J.}\ \bibnamefont {Plimpton}}, \bibinfo {author} {\bibfnamefont {P.~H.}\ \bibnamefont {Poole}},\ and\ \bibinfo {author} {\bibfnamefont {S.~C.}\ \bibnamefont {Glotzer}},\ }\href {https://doi.org/10.1103/physrevlett.80.2338} {\bibfield  {journal} {\bibinfo  {journal} {Physical Review Letters}\ }\textbf {\bibinfo {volume} {80}},\ \bibinfo {pages} {2338–2341} (\bibinfo {year} {1998})}\BibitemShut {NoStop}%
\bibitem [{\citenamefont {Morgan}(2020)}]{morgan2020argyrodite}%
  \BibitemOpen
  \bibfield  {author} {\bibinfo {author} {\bibfnamefont {B.}~\bibnamefont {Morgan}},\ }\href@noop {} {\bibinfo {title} {bjmorgan/data\_argyrodite\_disorder: Manuscript resubmission release}},\ \bibinfo {howpublished} {\url{https://zenodo.org/record/4338578}} (\bibinfo {year} {2020}),\ \bibinfo {note} {version 1.0, Zenodo, 17 Dec. 2020, \doi{10.5281/zenodo.4338578}}\BibitemShut {NoStop}%
\end{thebibliography}

\begin{thebibliography}{12}%
\makeatletter
\providecommand \@ifxundefined [1]{%
 \@ifx{#1\undefined}
}%
\providecommand \@ifnum [1]{%
 \ifnum #1\expandafter \@firstoftwo
 \else \expandafter \@secondoftwo
 \fi
}%
\providecommand \@ifx [1]{%
 \ifx #1\expandafter \@firstoftwo
 \else \expandafter \@secondoftwo
 \fi
}%
\providecommand \natexlab [1]{#1}%
\providecommand \enquote  [1]{``#1''}%
\providecommand \bibnamefont  [1]{#1}%
\providecommand \bibfnamefont [1]{#1}%
\providecommand \citenamefont [1]{#1}%
\providecommand \href@noop [0]{\@secondoftwo}%
\providecommand \href [0]{\begingroup \@sanitize@url \@href}%
\providecommand \@href[1]{\@@startlink{#1}\@@href}%
\providecommand \@@href[1]{\endgroup#1\@@endlink}%
\providecommand \@sanitize@url [0]{\catcode `\\12\catcode `\$12\catcode `\&12\catcode `\#12\catcode `\^12\catcode `\_12\catcode `\%12\relax}%
\providecommand \@@startlink[1]{}%
\providecommand \@@endlink[0]{}%
\providecommand \url  [0]{\begingroup\@sanitize@url \@url }%
\providecommand \@url [1]{\endgroup\@href {#1}{\urlprefix }}%
\providecommand \urlprefix  [0]{URL }%
\providecommand \Eprint [0]{\href }%
\providecommand \doibase [0]{https://doi.org/}%
\providecommand \selectlanguage [0]{\@gobble}%
\providecommand \bibinfo  [0]{\@secondoftwo}%
\providecommand \bibfield  [0]{\@secondoftwo}%
\providecommand \translation [1]{[#1]}%
\providecommand \BibitemOpen [0]{}%
\providecommand \bibitemStop [0]{}%
\providecommand \bibitemNoStop [0]{.\EOS\space}%
\providecommand \EOS [0]{\spacefactor3000\relax}%
\providecommand \BibitemShut  [1]{\csname bibitem#1\endcsname}%
\let\auto@bib@innerbib\@empty
\bibitem [{\citenamefont {Kapil}\ \emph {et~al.}(2022)\citenamefont {Kapil}, \citenamefont {Schran}, \citenamefont {Zen}, \citenamefont {Chen}, \citenamefont {Pickard},\ and\ \citenamefont {Michaelides}}]{Kapil_2022}%
  \BibitemOpen
  \bibfield  {author} {\bibinfo {author} {\bibfnamefont {V.}~\bibnamefont {Kapil}}, \bibinfo {author} {\bibfnamefont {C.}~\bibnamefont {Schran}}, \bibinfo {author} {\bibfnamefont {A.}~\bibnamefont {Zen}}, \bibinfo {author} {\bibfnamefont {J.}~\bibnamefont {Chen}}, \bibinfo {author} {\bibfnamefont {C.~J.}\ \bibnamefont {Pickard}},\ and\ \bibinfo {author} {\bibfnamefont {A.}~\bibnamefont {Michaelides}},\ }\href {https://doi.org/10.1038/s41586-022-05036-x} {\bibfield  {journal} {\bibinfo  {journal} {Nature}\ }\textbf {\bibinfo {volume} {609}},\ \bibinfo {pages} {512–516} (\bibinfo {year} {2022})}\BibitemShut {NoStop}%
\bibitem [{\citenamefont {Ravindra}\ \emph {et~al.}(2024)\citenamefont {Ravindra}, \citenamefont {Advincula}, \citenamefont {Shi}, \citenamefont {Coles}, \citenamefont {Michaelides},\ and\ \citenamefont {Kapil}}]{ravindra2024nuclearquantumeffectsinduce}%
  \BibitemOpen
  \bibfield  {author} {\bibinfo {author} {\bibfnamefont {P.}~\bibnamefont {Ravindra}}, \bibinfo {author} {\bibfnamefont {X.~R.}\ \bibnamefont {Advincula}}, \bibinfo {author} {\bibfnamefont {B.~X.}\ \bibnamefont {Shi}}, \bibinfo {author} {\bibfnamefont {S.~W.}\ \bibnamefont {Coles}}, \bibinfo {author} {\bibfnamefont {A.}~\bibnamefont {Michaelides}},\ and\ \bibinfo {author} {\bibfnamefont {V.}~\bibnamefont {Kapil}},\ }\href {https://arxiv.org/abs/2410.03272} {\bibfield  {journal} {\bibinfo  {journal} {arXiv preprint arXiv:2410.03272}\ } (\bibinfo {year} {2024})}\BibitemShut {NoStop}%
\bibitem [{\citenamefont {Batatia}\ \emph {et~al.}(2022)\citenamefont {Batatia}, \citenamefont {Kovacs}, \citenamefont {Simm}, \citenamefont {Ortner},\ and\ \citenamefont {Csanyi}}]{batatia2022mace}%
  \BibitemOpen
  \bibfield  {author} {\bibinfo {author} {\bibfnamefont {I.}~\bibnamefont {Batatia}}, \bibinfo {author} {\bibfnamefont {D.~P.}\ \bibnamefont {Kovacs}}, \bibinfo {author} {\bibfnamefont {G.~N.~C.}\ \bibnamefont {Simm}}, \bibinfo {author} {\bibfnamefont {C.}~\bibnamefont {Ortner}},\ and\ \bibinfo {author} {\bibfnamefont {G.}~\bibnamefont {Csanyi}},\ }in\ \href {https://openreview.net/forum?id=YPpSngE-ZU} {\emph {\bibinfo {booktitle} {Advances in Neural Information Processing Systems}}},\ \bibinfo {editor} {edited by\ \bibinfo {editor} {\bibfnamefont {A.~H.}\ \bibnamefont {Oh}}, \bibinfo {editor} {\bibfnamefont {A.}~\bibnamefont {Agarwal}}, \bibinfo {editor} {\bibfnamefont {D.}~\bibnamefont {Belgrave}},\ and\ \bibinfo {editor} {\bibfnamefont {K.}~\bibnamefont {Cho}}}\ (\bibinfo {year} {2022})\BibitemShut {NoStop}%
\bibitem [{\citenamefont {Hajibabaei}\ \emph {et~al.}(2025)\citenamefont {Hajibabaei}, \citenamefont {Baldwin}, \citenamefont {Csányi},\ and\ \citenamefont {Cox}}]{Hajibabaei_2025}%
  \BibitemOpen
  \bibfield  {author} {\bibinfo {author} {\bibfnamefont {A.}~\bibnamefont {Hajibabaei}}, \bibinfo {author} {\bibfnamefont {W.~J.}\ \bibnamefont {Baldwin}}, \bibinfo {author} {\bibfnamefont {G.}~\bibnamefont {Csányi}},\ and\ \bibinfo {author} {\bibfnamefont {S.~J.}\ \bibnamefont {Cox}},\ }\bibfield  {journal} {\bibinfo  {journal} {Physical Review Letters}\ }\textbf {\bibinfo {volume} {134}},\ \href {https://doi.org/10.1103/physrevlett.134.026306} {10.1103/physrevlett.134.026306} (\bibinfo {year} {2025})\BibitemShut {NoStop}%
\bibitem [{\citenamefont {Müller}\ \emph {et~al.}(2021)\citenamefont {Müller}, \citenamefont {Ertural}, \citenamefont {Hempelmann},\ and\ \citenamefont {Dronskowski}}]{M_ller_2021}%
  \BibitemOpen
  \bibfield  {author} {\bibinfo {author} {\bibfnamefont {P.~C.}\ \bibnamefont {Müller}}, \bibinfo {author} {\bibfnamefont {C.}~\bibnamefont {Ertural}}, \bibinfo {author} {\bibfnamefont {J.}~\bibnamefont {Hempelmann}},\ and\ \bibinfo {author} {\bibfnamefont {R.}~\bibnamefont {Dronskowski}},\ }\href {https://doi.org/10.1021/acs.jpcc.1c00718} {\bibfield  {journal} {\bibinfo  {journal} {J. Phys. Chem. C}\ }\textbf {\bibinfo {volume} {125}},\ \bibinfo {pages} {7959} (\bibinfo {year} {2021})}\BibitemShut {NoStop}%
\bibitem [{\citenamefont {Kreuer}(1996)}]{Kreuer_1996}%
  \BibitemOpen
  \bibfield  {author} {\bibinfo {author} {\bibfnamefont {K.-D.}\ \bibnamefont {Kreuer}},\ }\href {https://doi.org/10.1021/cm950192a} {\bibfield  {journal} {\bibinfo  {journal} {Chemistry of Materials}\ }\textbf {\bibinfo {volume} {8}},\ \bibinfo {pages} {610–641} (\bibinfo {year} {1996})}\BibitemShut {NoStop}%
\bibitem [{\citenamefont {Di~Pino}\ \emph {et~al.}(2023)\citenamefont {Di~Pino}, \citenamefont {Perez~Sirkin}, \citenamefont {Morzan}, \citenamefont {Sánchez}, \citenamefont {Hassanali},\ and\ \citenamefont {Scherlis}}]{Di_Pino_2023}%
  \BibitemOpen
  \bibfield  {author} {\bibinfo {author} {\bibfnamefont {S.}~\bibnamefont {Di~Pino}}, \bibinfo {author} {\bibfnamefont {Y.~A.}\ \bibnamefont {Perez~Sirkin}}, \bibinfo {author} {\bibfnamefont {U.~N.}\ \bibnamefont {Morzan}}, \bibinfo {author} {\bibfnamefont {V.~M.}\ \bibnamefont {Sánchez}}, \bibinfo {author} {\bibfnamefont {A.}~\bibnamefont {Hassanali}},\ and\ \bibinfo {author} {\bibfnamefont {D.~A.}\ \bibnamefont {Scherlis}},\ }\bibfield  {journal} {\bibinfo  {journal} {Angewandte Chemie International Edition}\ }\textbf {\bibinfo {volume} {62}},\ \href {https://doi.org/10.1002/anie.202306526} {10.1002/anie.202306526} (\bibinfo {year} {2023})\BibitemShut {NoStop}%
\bibitem [{\citenamefont {Muñoz-Santiburcio}\ and\ \citenamefont {Marx}(2021)}]{Mu_oz_Santiburcio_2021}%
  \BibitemOpen
  \bibfield  {author} {\bibinfo {author} {\bibfnamefont {D.}~\bibnamefont {Muñoz-Santiburcio}}\ and\ \bibinfo {author} {\bibfnamefont {D.}~\bibnamefont {Marx}},\ }\href {https://doi.org/10.1021/acs.chemrev.0c01292} {\bibfield  {journal} {\bibinfo  {journal} {Chemical Reviews}\ }\textbf {\bibinfo {volume} {121}},\ \bibinfo {pages} {6293–6320} (\bibinfo {year} {2021})}\BibitemShut {NoStop}%
\bibitem [{\citenamefont {Cheng}\ \emph {et~al.}(2021)\citenamefont {Cheng}, \citenamefont {Bethkenhagen}, \citenamefont {Pickard},\ and\ \citenamefont {Hamel}}]{Cheng_2021}%
  \BibitemOpen
  \bibfield  {author} {\bibinfo {author} {\bibfnamefont {B.}~\bibnamefont {Cheng}}, \bibinfo {author} {\bibfnamefont {M.}~\bibnamefont {Bethkenhagen}}, \bibinfo {author} {\bibfnamefont {C.~J.}\ \bibnamefont {Pickard}},\ and\ \bibinfo {author} {\bibfnamefont {S.}~\bibnamefont {Hamel}},\ }\href {https://doi.org/10.1038/s41567-021-01334-9} {\bibfield  {journal} {\bibinfo  {journal} {Nature Physics}\ }\textbf {\bibinfo {volume} {17}},\ \bibinfo {pages} {1228–1232} (\bibinfo {year} {2021})}\BibitemShut {NoStop}%
\bibitem [{\citenamefont {Reinhardt}\ \emph {et~al.}(2022)\citenamefont {Reinhardt}, \citenamefont {Bethkenhagen}, \citenamefont {Coppari}, \citenamefont {Millot}, \citenamefont {Hamel},\ and\ \citenamefont {Cheng}}]{Reinhardt_2022}%
  \BibitemOpen
  \bibfield  {author} {\bibinfo {author} {\bibfnamefont {A.}~\bibnamefont {Reinhardt}}, \bibinfo {author} {\bibfnamefont {M.}~\bibnamefont {Bethkenhagen}}, \bibinfo {author} {\bibfnamefont {F.}~\bibnamefont {Coppari}}, \bibinfo {author} {\bibfnamefont {M.}~\bibnamefont {Millot}}, \bibinfo {author} {\bibfnamefont {S.}~\bibnamefont {Hamel}},\ and\ \bibinfo {author} {\bibfnamefont {B.}~\bibnamefont {Cheng}},\ }\href {https://doi.org/10.1038/s41467-022-32374-1} {\bibfield  {journal} {\bibinfo  {journal} {Nat. Commun.}\ }\textbf {\bibinfo {volume} {13}},\ \bibinfo {pages} {4707} (\bibinfo {year} {2022})}\BibitemShut {NoStop}%
\bibitem [{\citenamefont {Donati}\ \emph {et~al.}(1998)\citenamefont {Donati}, \citenamefont {Douglas}, \citenamefont {Kob}, \citenamefont {Plimpton}, \citenamefont {Poole},\ and\ \citenamefont {Glotzer}}]{Donati_1998}%
  \BibitemOpen
  \bibfield  {author} {\bibinfo {author} {\bibfnamefont {C.}~\bibnamefont {Donati}}, \bibinfo {author} {\bibfnamefont {J.~F.}\ \bibnamefont {Douglas}}, \bibinfo {author} {\bibfnamefont {W.}~\bibnamefont {Kob}}, \bibinfo {author} {\bibfnamefont {S.~J.}\ \bibnamefont {Plimpton}}, \bibinfo {author} {\bibfnamefont {P.~H.}\ \bibnamefont {Poole}},\ and\ \bibinfo {author} {\bibfnamefont {S.~C.}\ \bibnamefont {Glotzer}},\ }\href {https://doi.org/10.1103/physrevlett.80.2338} {\bibfield  {journal} {\bibinfo  {journal} {Physical Review Letters}\ }\textbf {\bibinfo {volume} {80}},\ \bibinfo {pages} {2338–2341} (\bibinfo {year} {1998})}\BibitemShut {NoStop}%
\bibitem [{\citenamefont {Morgan}(2021)}]{Morgan_2021}%
  \BibitemOpen
  \bibfield  {author} {\bibinfo {author} {\bibfnamefont {B.~J.}\ \bibnamefont {Morgan}},\ }\href {https://doi.org/10.1021/acs.chemmater.0c03738} {\bibfield  {journal} {\bibinfo  {journal} {Chemistry of Materials}\ }\textbf {\bibinfo {volume} {33}},\ \bibinfo {pages} {2004–2018} (\bibinfo {year} {2021})}\BibitemShut {NoStop}%
\end{thebibliography}

\end{document}